\documentclass[printer]{aa}  

\usepackage{graphicx}
\usepackage{subfig}
\usepackage{float}
\usepackage{xcolor}
\usepackage[inline]{enumitem}
\usepackage{txfonts}
\newcommand{\myparagraph}[1]{\paragraph{#1}\mbox{}\\[0.3cm]}

\newcommand{\emsta}[0]{EM-$\{\}$}
\newcommand{\emrho}[0]{EM-$\{\rho\}$}
\newcommand{\emrhoT}[0]{EM-$\{\rho,T\}$}

\begin{document} 
\bibpunct{(}{)}{;}{a}{}{,}

\title{Satellite dwarf galaxies: Stripped but not quenched.}

\author{Loic Hausammann
  \inst{1}
  \and
  Yves Revaz\inst{1}
  \and
  Pascale Jablonka\inst{1,2}
}

\institute{Laboratoire d'Astrophysique, Ecole Polytechnique F\'ed\'erale de Lausanne (EPFL), 1290 Sauvergny, Switzerland\\
  \email{loic.hausammann@epfl.ch}
  \and
   CNRS UMR 8111,  GEPI, Observatoire de Paris, Université PSL, 92125 Meudon Cedex, France\\}

\date{Received September XXX; accepted XXX}

% \abstract{}{}{}{}{} 
% 5 {} token are mandatory

\abstract{In the Local Group, quenched gas-poor dwarfs galaxies are most often found
  close to the Milky Way and Andromeda, while star forming gas-rich ones are
  located at greater distances. This so-called morphology-density relation is
  often interpreted as the consequence of the ram pressure stripping of the
  satellites during their interaction with the Milky Way hot halo gas.  While
  this process has been often investigated, self-consistent high resolution
  simulations were still missing. In this study, we have analysed the impact of
  both the ram pressure and tidal forces induced by a host galaxy on
  dwarf models as realistic as possible emerging from cosmological simulations.
  These models were re-simulated using both a wind tunnel and a moving box
  technique.  The secular mass growth of the central host galaxy, as well as the
  gas density and temperature profiles of its hot halo have been taken into
  account.  We show that while ram pressure is very efficient at stripping the
  hot and diffuse gas of the dwarf galaxies, it can remove their cold gas ($T < 10^3$~[K]) only
  in very specific conditions. Depending on the infall time of the satellites
  relatively to the build-up stage of the massive host, star formation can thus be
  prolonged instead of being quenched. This is the direct consequence of the clumpy 
  nature of the cold gas and the thermal pressure the hot gas exerts onto it. 
  %We discuss
  %the possibility that the Milky Way had an unusual accretion history and
  %suggest possible further numerical improvements.
  %Our results shed a new light on the variety of satellite populations, when accounting for
  %the recently observed dwarfs around Milky Way-like analogues
  We discuss the possibility that the variety in satellite populations among Milky Way-like galaxies 
  reflects their accretion histories.
}


   \maketitle
   %------------------------------------------------------------
   %                          INTRO
   %------------------------------------------------------------

   %------------------------------------------------------------%
   %------------------------- Section --------------------------%
   %------------------------------------------------------------%
   \section{Introduction}
   % PASCALE: DR2 Gaia mesure mvt propre -> impact?
   % PASCALE: Question impact MW on satellites claimed to be solved. Qu est ce qui a change?

   % introduce dwarf galaxies
   
   Dwarf galaxies are the faintest galaxies found in the Universe. In
   a hierarchical $\Lambda$CDM framework, they are the most common
   systems and, in their early evolution phase, they can serve as building
   blocks of larger galaxies. Suggestions are made that dwarfs could have
   played a substantial role during the epoch of reionization
   \citep{atek_are_2015,robertson_cosmic_2015,bouwens_uv_2015}. 
   Understanding their role in this context requires a detailed
   picture of their formation and evolution.
   % which requires accurate numerical modelization.
   
   Noteworthily, dwarf galaxies have challenged $\Lambda$CDM on a
   number of questions, such as the missing satellites
   \citep{moore_dark_1999,klypin_where_1999}, the too-big-to-fail
   \citep{boylan-kolchin_too_2011,boylan-kolchin_milky_2012} or the
   core-cusp
   \citep{navarro_structure_1996,navarro_universal_1997,moore_evidence_1994}
   problems \citep[see][for a complete review]{bullock_small-scale_2017}.
   These issues were originally highlighted for dark matter only cosmological simulations.
   However, since these pioneering simulations, major improvements have been achieved, in particular thanks
   to the inclusion of the evolution of the baryons in the simulations, but also thanks to very significant progresses
   in numerical methods \citep{springel_cosmological_2005,wiersma_chemical_2009,aubert_reionization_2010,hahn_multi-scale_2011,
     durier_implementation_2012,haardt_radiative_2012,hopkins_general_2013,revaz_computational_2016}.
   As a consequence, when baryonic physics is properly included, the numerical simulations are now able to reproduce a large variety of observed properties
   \citep{valcke_simulations_2008,revaz_dynamical_2009,sawala_formation_2010,schroyen_simulations_2011,revaz_dynamical_2012,cloet-osselaer_degeneracy_2012,sawala_local_2012,
     cloet-osselaer_numerical_2014,sawala_apostle_2016,wetzel_reconciling_2016,fitts_fire_2017,maccio_edge_2017,escala_modelling_2018,revaz_pushing_2018}.
   High resolution cosmological hydro-dynamical simulations of the Local Group such as APOSTLE \citep{sawala_apostle_2016} or Latte
   \citep{wetzel_reconciling_2016,garrison-kimmel_local_2018} also lead to solving the cosmological problems previously mentioned.
   However a global consensus on whether or not those problems are definitely solved is still missing.
   See \citet{bullock_small-scale_2017} for a review.

   %reproduce mophology relation

   % interaction + ram/tidal

   While a proper treatment of the intrinsic evolution of the dwarf
   galaxies is mandatory, the possible impact of the environment of these systems ought to
   be understood as well. Observations have indeed highlighted a
   morphology-density relation in the Local Group
   \citep{einasto_missing_1974,mcconnachie_observed_2012}. Gas-deficient
   galaxies are preferentially found close to either the Milky Way or
   M31, while gas-rich dwarfs are found at larger galacto-centric
   distances. 
   This relation could result from the interaction between satellite systems
   and their massive host, through both tidal and ram pressure stripping.
   While tidal stripping is a pure gravitational process, ram
   pressure stripping is an hydrodynamical one, resulting from the
   interaction between the interstellar medium (ISM) of the dwarf and
   the hot virialized diffused gas of its host galaxy, that can reach
   temperature up to $\sim 10^6\,\rm{K}$, for a Milky Way analogue.
   The stripping of the dwarf galaxy results from a momentum exchange
   between the two gas components. 

   Ram pressure, with or without the help of tidal stripping has also been
   mentioned to possibly solve the missing satellites problem
   \citep{del_popolo_small_2017,arraki_effects_2014}. Indeed, the
   quick removal of the ISM of the dwarf makes its luminosity drop
   down to the point of hampering its detection.  The dynamics of the
   dwarf is also modified, impacting its mass distribution, eventually
   turning a cuspy profile into a cored one.  While
   \citet{mayer_simultaneous_2006} and \citet{simpson_quenching_2018}
   found that ram pressure and tidal stripping are efficient at
   removing the gas of the dwarf galaxies and at quenching their star
   formation, others, such as 
   \citet{emerick_gas_2016} and \citet{wright_reignition_2019} found it far less
   so and sometimes even able to slightly enhance star formation.  While most
   of those studies reproduce the relation between the dwarf neutral
   gas (HI) fraction and their distance to the host galaxy
   \citep{grcevich_hi_2010}, some are not run in a cosmological
   context and the treatment of the baryonic physics is generally
   incomplete. For example, hydrogen self-shielding against UV-ionizing photons, 
   that let the gas efficiently cool below $10^4\,\rm{K}$ 
   is missing. This hampers the capturing of the multi-phase
   structure of the dense star forming gas.
   
   The present work is based on the high resolution zoom-in cosmological
   simulations of \citet{revaz_pushing_2018}. A
   volume of $(3.4\,\rm{Mpc/h})^3$ has served the analysis of dwarf
   galaxies outside the influence of a massive Milky-Way like
   galaxy. It was shown that, when baryonic physics and
   UV-background is included,
   in vast majority, the observed variety
   of galaxy properties, star formation histories, metallicity
   distribution, stellar chemical abundance ratios, kinematics, and
   gas content, was reproduced in detail as a natural consequence of
   the $\Lambda$CDM hierarchical formation sequence.
   Some systems though could not be
   adequately reproduced, such as the Fornax dwarf spheroidal galaxy
   (dSph), which is dominated by an intermediate stellar population
   \citep{boer_star_2012}, or the Carina dSph
   \citep{de_boer_episodic_2014}, which exhibits very distinct
   peaks of star formation. Others such Leo P or Leo T
   \citep{mcquinn_leo_2015,weisz_star_2012} have more extended star
   formation histories than can be predicted as the result of their
   low halo mass and the impact of the UV-background heating.

   The question of when and how the Milky-Way, or similar central host
   galaxy, can impact the evolution of its satellites is at the heart
   of this study. This can also shed light on the origin of the above
   mentioned Local Group dSphs, which stand as exceptions of a general
   framework. To this end, we extracted a series of models from
   \citet{revaz_pushing_2018} and re-simulated them by taking into
   account a Milky Way-like environment.  Two sets of simulations are
   presented in the following: a wind tunnel, which investigates the
   impact of the ram pressure alone and a moving box, which includes
   the tidal forces as well.

   The structure of this paper is the following. In
   Section~\ref{sec:tools}, we present our numerical tools, the code
   \texttt{GEAR}, the wind tunnel and the moving box techniques.  In
   Section \ref{sec:model} we describe the initial conditions of our
   dwarf models as well as their orbits.  The different Milky Way
   models are also presented.  In Section \ref{sec:sim_set} the sets of runs
   for our two different simulation techniques are detailed. 
   Our results are presented in Section \ref{sec:results} and a
   discussion is proposed in Section~\ref{sec:discussion}, followed by
   a short conclusion in Section~\ref{sec:conclusion}.

   %------------------------------------------------------------%
   %------------------------- Section --------------------------%
   %------------------------------------------------------------%
   \section{Numerical tools}\label{sec:tools}

   Our simulations involve two galaxies: the satellite, a dwarf galaxy and its host, a Milky Way-like galaxy.   
   The dwarf galaxy is self-consistently simulated as an N-body system using the code \texttt{GEAR}.
   To capture the ram pressure induced by the hot host halo, we used a wind tunnel method where
   gas particles are injected and interact with the dwarf galaxy.
   The effect of tidal forces is included by extending the wind tunnel simulation with a moving box technique.
   There, the gravity of the host galaxy is modelled by a potential that may evolve with time.
   Those different techniques are succinctly presented in this section.

   \subsection{\texttt{GEAR}}\label{sec:gear}
   \texttt{GEAR} is a chemo-dynamical Tree/SPH code based on \texttt{GADGET-2} \citep{springel_cosmological_2005}. Its original version 
   was described in \citet{revaz_dynamical_2012} with some improvements discussed in \citet{revaz_computational_2016}
   and \citet{revaz_pushing_2018}.
   Gas radiative cooling and UV-background heating are computed through the \texttt{GRACKLE} library \citep{smith_grackle:_2017}, using its equilibrium mode.
   In this mode, the cooling due to the primordial elements are precomputed following the assumption of ionization equilibrium under the presence of a photoionizing 
   UV-background \citep{haardt_radiative_2012}.
   Cooling from metals is included using a simple method where predictions for a solar-metalicity gas computed from the \texttt{CLOUDY} code \citep{ferland_2017_2017} 
   are scaled according to the gas metallicity \citep[see][for the details of the method]{smith_grackle:_2017}.
   The cooling due to the $H_2$ molecule is not included.
   Hydrogen self-shielding is included by suppressing the UV-background heating for densities above $0.007$ cm$^{-3}$ \citep{aubert_reionization_2010}.
   A lower temperature limit of 10 [K] is imposed.

   Star formation is performed using a modified version of the Jeans pressure \citep{hopkins_self-regulated_2011} and an efficiency $c_\star = 0.01$.
   The chemical evolution scheme includes Type Ia and II supernova with yields from \citet{kobayashi_history_2000} and \citet{tsujimoto_relative_1995} respectively.
   Exploding supernovae are computed stochastically using a random discrete IMF sampling (RIMFS) scheme \citep{revaz_computational_2016}.
   An energy of $10^{50}$erg is released per supernova into the ISM, following the thermal blastwave-like feedback scheme \citep{stinson_star_2006}.
   We used the smooth metalicity scheme \citep{okamoto_effects_2005,tornatore_chemical_2007,wiersma_chemical_2009} to further mix the polluted gas.
   Stellar V-band luminosities are computed using \citet{vazdekis_new_1996} relations and our initial mass function (IMF) is the revised IMF of \citet{kroupa_variation_2001}.
   \texttt{GEAR} includes individual and adaptive time steps \citep{durier_implementation_2012} and the pressure-entropy SPH formulation \citep{hopkins_general_2013}
   which ensures the correct treatment of fluid mixing instabilities, essential in the RPS simulations.

   In the present study, the physical models and its parameters are identical to the one used in \citet{revaz_pushing_2018}, where the properties of a few Local Group's dwarf galaxy such as NGC 6622, 
   Andromeda II, Sculptor and Sextans have been reproduced in great details.

   \subsection{Wind tunnel}
   In order to study RP stripping, we supplement \texttt{GEAR} with a wind tunnel setup.
   A wind tunnel simulation consists in an object  (an isolated galaxy in our case), placed in a box in which gas particles, called hereafter wind particles, are injected from one side (the front) and removed from the opposite one (the back).
   In-between wind particles may interact with the object and in particular with its gaseous component.
   In our implementation, the behaviour of particles at the box side, meaning, the six box faces different from the front and back ones
   differ according to their origin.     
   If particles are gas from the wind, we apply periodic boundaries. On the contrary, if particles where gas, initially belonging to the satellite, 
   they are removed. 
   Finally, we remove all type of particles that cross the front side with negative velocities, that is moving against the wind.

   The details of the parameters explored through those wind tunnel simulations will be presented in Section~\ref{sec:model}.
   While being the perfect tool to study RP and in particular the effect of a variation of the wind density, temperature and velocity,
   wind tunnels simulations do no include any tidal effect and its dependence along the satellite orbit.

   \subsection{Moving box}\label{sec:implementation}

   We complemented the wind tunnels simulations with moving box simulations.
   This simulation technique introduced by \citet{nichols_post-infall_2015} allows to add the tidal stripping a satellite may suffer along its orbit, while ensuring
   simulations to run with the same very high resolution.
   Hereafter, we  present a brief summary of this methods, including minor updates.

   The moving box consists in a wind tunnel simulation supplemented with the gravitational forces between the host (a fixed potential) and a satellite moving along its orbit.
   Instead of launching a satellite in an orbit around a host potential,   
   the satellite is placed inside a non inertial box corresponding to a frame in motion around the host potential. 
   In addition to its motion along the orbit, we supplement the box with a rotation motion in order to keep the particles injection on the same front side. The latter is simulated by
   implementing fictitious forces induced by both the rotation and orbital motion of the box.
   This method is a CPU-economic way of simulating what a galaxy would experiment while orbiting around its host without the necessity to include
   the entire hot gas halo that would requires important memory and CPU resources.
   
   Stars and dark matter are not  sensitive to the hydrodynamical forces. However, they are indirectly affected by the RP through 
   the gravitational restoring force the RP stripped gas will exerts on both of them (see the parachute effect described in \citet{nichols_post-infall_2015}).
   This indirect interaction is responsible for a continuous drift of the satellite with respect to the box centre, which, in extreme case could make it leave the box.
   To avoid this, we apply an ad hoc correcting force which depends on the centre of the dwarf, defined as the  
   centre of mass of the 64 star and dark matter particles of the dwarf having the lowest total specific energy.
   This definition is sensitively optimized compared to the one performed by \citet{nichols_post-infall_2015}, where only the potential
   energy was used, leading to the impossible differentiation between bounded particle and particles passing through at high velocity. 
   Once the dwarf centre is defined, an harmonic force is apply to all particles, where the magnitude of the force scales with the distance between its centre and the centre of the box.
   The impact of this procedure on the satellite orbit is small. Only a slight reduction of the apocentre (about 15\%) as well as of the velocity at pericentre (about 10\%) after $10\,\rm{Gyr}$ is observed, with respect to 
   the expected theoretical orbit where a satellite is considered as a point mass.   
   One restriction of the method is the ill defined behaviour of the wind particles creation when the host centre lie inside the simulation box.
   Indeed, in the case where the host centre would enter the box, there is no way to clearly define a front face where we could inject the wind particles.   
   Therefore we restrained the orbits to radius larger than the box size. 
   The details of the orbits as well as the set of simulations performed are described in Section~\ref{sec:model}.

   %------------------------------------------------------------%
   %------------------------- Section --------------------------%
   %------------------------------------------------------------%

   \section{Models}\label{sec:model}
    
   \subsection{Dwarf models} \label{sec:dwarfs}
   
   All our dwarf models have been extracted from the cosmological zoom-in simulations published in \citet{revaz_pushing_2018}. We refer to 
   this paper regarding the name of dwarf models. 
   27 dwarfs have been simulated from $z_\textrm{init} = 70$ until $z=0$, assuming \citet{planck_collaboration_planck_2015} cosmological parameters, 
   with a gravitational softening of $10$ and $50\,\rm{pc/h}$ for the gas and dark matter respectively and a mass resolution of $1'024$ M$_\odot$/$h$ for the stellar, $4'096 $ M$_\odot$/h for the gas and $22'462$ M$_\odot$/h for the dark matter.
   Despite having still an important gas component at the injection redshift, none of the simulated dwarf show a disky structure.
   This is due to the lack of angular momentum accretion as well as the strong stellar feedback that continuously heats gas,
   maintaining it in a spherical structure around the dwarf.
      
   In a first step, in order to test the ram pressure under a large number of parameters at low computational cost, we mainly focused on model \texttt{h159} in our
   wind tunnel simulations. This model is a quenched galaxy dominated by an old stellar population with a final V-band luminosity of $0.42 \cdot 10^6\,\rm{L_\odot}$, a virial mass of
   $M_\textrm{200} = 5.41 \cdot 10^8$ M$_\odot$ (See Table 1. of \citet{revaz_pushing_2018}). 
   Because of its low stellar mass and quenched star formation history this model is quickly simulated over one Hubble time.
   While results presented in Sec.~\ref{sec:windtunnel_results} only rely on this galaxy, it is worth noting that similar results have been obtained with six more massive galaxies 
   (see Table~\ref{tab:addwt}).
      
   In a second step, in our moving box simulations, seven galaxies have been selected according to their star formation history, spanning a total halo mass in the range $M_\textrm{200} = 5.4$ to $26.2 \times 10^8$ M$_\odot$ (see Table~\ref{tab:addmb}). In Sec.~\ref{sec:movingbox_results}, we focus on the two most representative cases, \texttt{h070} and \texttt{h159}. 
   Model \texttt{h070} is brighter than model \texttt{h159} with an extended star formation history. It perfectly reproduces the observed properties of the
   Sculptor dSph.

   \label{sec:extraction}
   Each selected dwarf model has been extracted from the cosmological simulation at $z_\textrm{ext}=2.4$ and converted from 
   comoving coordinates to physical ones. The extraction radius is taken as the virial radius $R_\textrm{200}$, where $R_\textrm{200}$ is the radius of a sphere that contains
   a mean mass density equal to 200 times the critical density of the Universe. For a dwarf spheroidal galaxy in a $\Lambda$CDM Universe, $R_\textrm{200}$ is of the order of $30\,\rm{kpc}$, 
   much larger than the stellar component ($\sim 1\,\rm{kpc}$).
   Using $R_\textrm{200}$ has the advantage of being large enough to minimize perturbation due to the extraction and small enough to keep a reasonable box size.
   We tested our extraction method and how it can perturb the evolution of the dwarf 
   by comparing the cumulative number of stars formed between the initial cosmological simulation and the extracted one at $z=0$.
   The perturbation has been found to be negligible, of the order of a perturbation induced by changing the random number seed.
   Simulating the late stage of dwarf galaxies out of a full cosmological context is justified by their merger history 
   \citep{revaz_dynamical_2012,fitts_no_2018,cloet-osselaer_numerical_2014} that finish early enough ($z \approx 5$ in our simulations) to be almost isolated for most of its life. 
   
   We chose the extraction redshift $z_\textrm{ext}$ on the following basis. 
   Due to the mergers at high redshift, $z_\textrm{ext}$ must be low enough to avoid a perturbation from a major merger (mass ratio of 0.1 in \citet{fitts_no_2018}).
   It must be high enough to ensure the quenched dwarfs to be still star forming ($t \lesssim 2,\rm{Gyr}$ for the faintest models like \texttt{h159}) in order to study the MW perturbation on its star formation history. We therefore choose $z_\textrm{ext}=2.4$.
   This choice corresponds to a satellite infall time of about 9 Gyr ago, considered as an early infall time according to \citep{wetzel_satellite_2015}. 
   A rather high fraction of present satellite galaxies, 15.8\%, have approximately this first infall time \citep{simpson_quenching_2018}.

   \subsubsection{Milky Way models at $z=0$}\label{sec:milky_way}
   The Milky Way mass model at $z=0$ is composed of two Plummer profiles representing a bulge and a disk, and an NFW profile representing its dark halo.
   The adopted parameters for these three components are given in Tab.~\ref{tab:milky_way} and are similar to the ones used in \citet{nichols_post-infall_2015}.

   \begin{table*}
     \begin{center}
	   \caption{Milky Way model parameters used at $z=0$ \citep{nichols_post-infall_2015}.
             The analytic potential of each component is provided in the first column ($\phi(R)$) along with its parameters in the second column.
             The last column provides the corresponding references.\label{tab:milky_way}}
	   \begin{tabular}{cccc}
	   \hline\hline
	   & $\phi(R)$ & Parameters & Reference\\ \hline
	   Bulge & $-GM/\sqrt{R^2 + a^2}$ & $\begin{array}{c}M = 1.3\cdot 10^{10}M_\odot \\ a = 0.5 \text{kpc}\end{array}$ & \citet{xue_milky_2008} \\
	   Disk & $-GM/\sqrt{R^2 + a^2}$ & $\begin{array}{c}M = 5.8\cdot 10^{10}M_\odot \\ a = 5 \text{kpc}\end{array}$ & \citet{xue_milky_2008} \\
	   Halo & $-GM_\textrm{vir}\ln(1+cR/R_\textrm{vir})/[\ln(1+c)-c/(1+c)]$ & $\begin{array}{c}M_\textrm{vir} = 8\cdot 10^{11}M_\odot \\ c = 21 \\ R_\textrm{vir} = 240 \text{kpc}\end{array}$ &     \citet{kafle_shoulders_2014} \\ \hline\hline
	   \end{tabular}
	 \end{center}
   \end{table*}

\begin{figure}
	   \begin{center}
		   \includegraphics[width=0.5\textwidth]{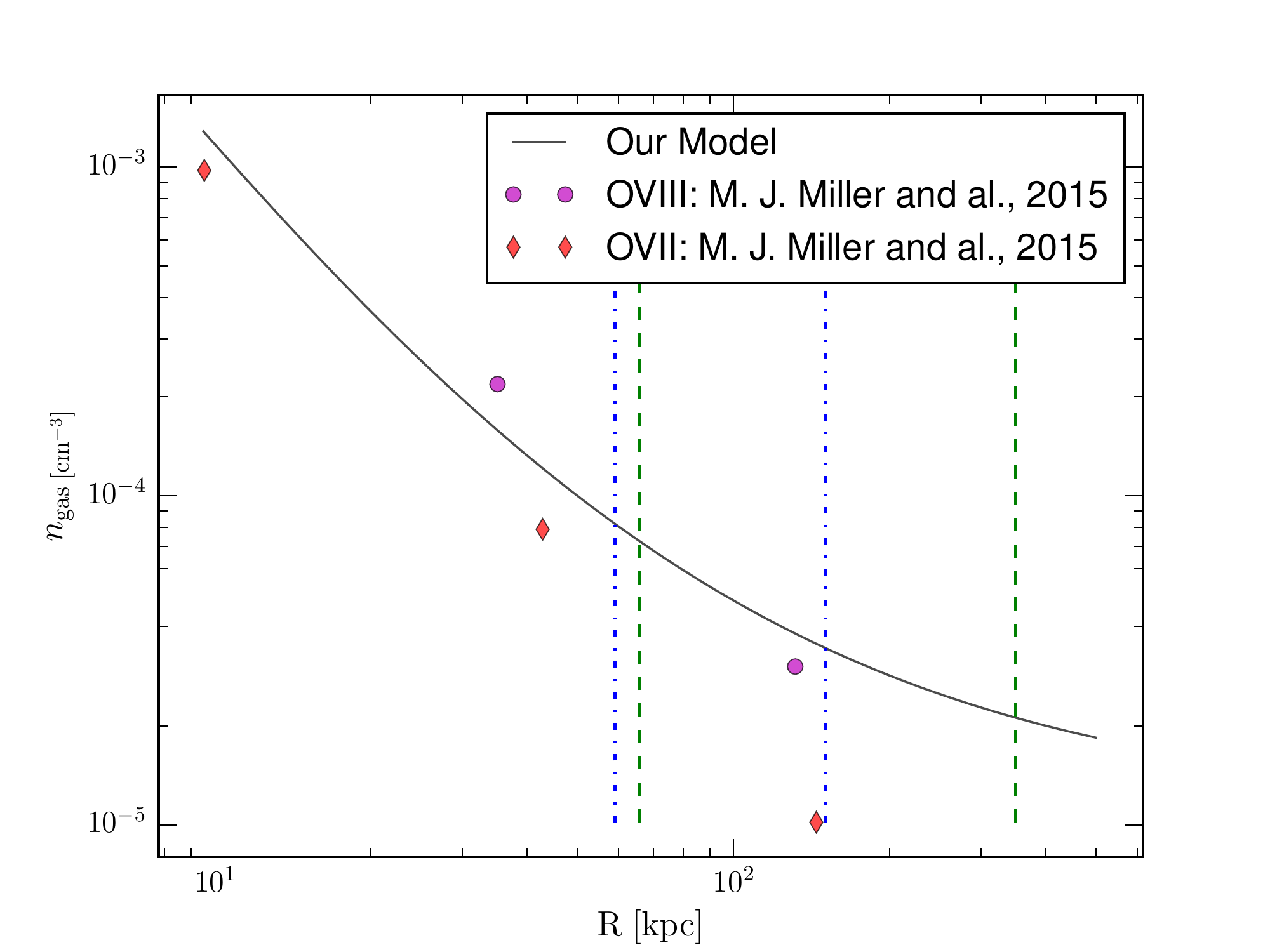}
		   \caption{Gas density model of the MilkyWay's hot halo (black line) compared to observational data from \citet{miller_constraining_2015}.
		     The two vertical lines correspond to the minimal pericentre and maximal apocentre of the satellite orbits explored in this work (blue static potential and green evolving potential).
		   \label{fig:hot_halo}}
	   \end{center}
   \end{figure}

   The gas density of the hot halo is computed by assuming the hydrostatic equilibrium of an ideal isothermal gas of hydrogen and helium. Formally the total gas density profile $\rho(r)$ 
   or equivalently the electron density profile $n_\textrm{e}$ is obtained by solving :
   \begin{equation}
   \frac{n_\textrm{e}(r)}{n_\textrm{e,0}} = \frac{\rho(r)}{\rho_0} = \exp\left( - \frac{\mu m_\textrm{p}}{k_\textrm{B} T} \left[ \phi(r) - \phi_0 \right] \right),
   \end{equation}
   where, $T$ is the constant gas temperature, $\phi$ the total potential, $\mu$ the mean molecular weight, $m_\textrm{p}$ the proton mass and $k_\textrm{B}$ the Boltzmann constant.
   $n_\textrm{e,0}$, $\rho_0$ and $\phi_0$ are respectively the electron density, total gas density and potential at the centre of the galaxy.
   Following \citet{nichols_post-infall_2015}, we fixed $n_\textrm{e,0}$ to $2\cdot 10^{-4}$ cm$^{-3}$ at $50$~kpc.
   The resulting density profile is displayed in Fig.~\ref{fig:hot_halo} and compared to the data of \citet{miller_constraining_2015}.
   The observed density and temperature intervals are $\rho \in [10^{-5}, 10^{-2} ]$ atom/cm$^{3}$ and $T \in [ 1.5 \cdot 10^6, 3 \cdot 10^6]$ K at radii smaller than 100 kpc and are consistent 
   with our MW model.
   At large radii our model slightly over-predicts the density. This is however unimportant as in any case, the ram pressure will not be negligible at those large radius 
   compared to smaller ones.
   The two vertical lines shown on Fig.~\ref{fig:hot_halo} indicate the minimal pericentre and maximal apocentre of the satellite orbits explored in this work and give and idea of
   the density studied in this work.

   \subsubsection{Time evolution of the models}

   All along a Hubble time, a  Milky Way-like galaxy see its mass growing through a succession
   of merger and accretion events. This mass grows and subsequently the increase of its gas halo and in particular its temperature
   through thermalisation has potentially a strong impact on the ram pressure and tidal tripping of its dwarf satellites.
   For this purpose, we considered the mass evolution of the MW model by defining three different evolution modes (EM). In all of them,
   the MW ends up with the same properties at $z=0$: 
  \begin{itemize}
  	\item    Static (\emsta): The MW does not evolve: Its potential remains fixed, equal to the one defined at $z=0$.
  	Similarly, the density and temperature of the halo gas stay constant. 
  	\item    Dynamic with a constant temperature (\emrho): The MW potential evolves through an increase of its total mass and size,
  	together with the density of the hot component. The temperature of the gas is however kept fixed.
  	\item    Dynamic with a dynamic temperature (\emrhoT): In addition to the second mode the gas temperature evolves too.
  \end{itemize}

   \subsubsection{Mass and size evolution}
   
   Figure~\ref{fig:evolving_milky_way} displays the time evolution of the mass and size of our MW model used in 
   the evolution mode \emrho  \,and \emrhoT.
   Those curves are computed from the model Louise of the ELVIS simulations \citep{garrison-kimmel_elvis:_2014}, 
   where the mass growth of several simulated galaxies is studied.
   The mass and size of the Louise galaxy is scaled in order to match exactly our non-evolving Milky Way model at $z=0$. 
   As in  \citet{garrison-kimmel_elvis:_2014} the Louise galaxy  was fitted using an NFW profile, we use the scale radius 
   ($R_s = R_\textrm{vir}/c$) as a scaling for the Plummer softening parameter $a$.

   \subsubsection{Temperature evolution}
   
\begin{figure}
     \begin{center}
       \includegraphics[width=0.5\textwidth]{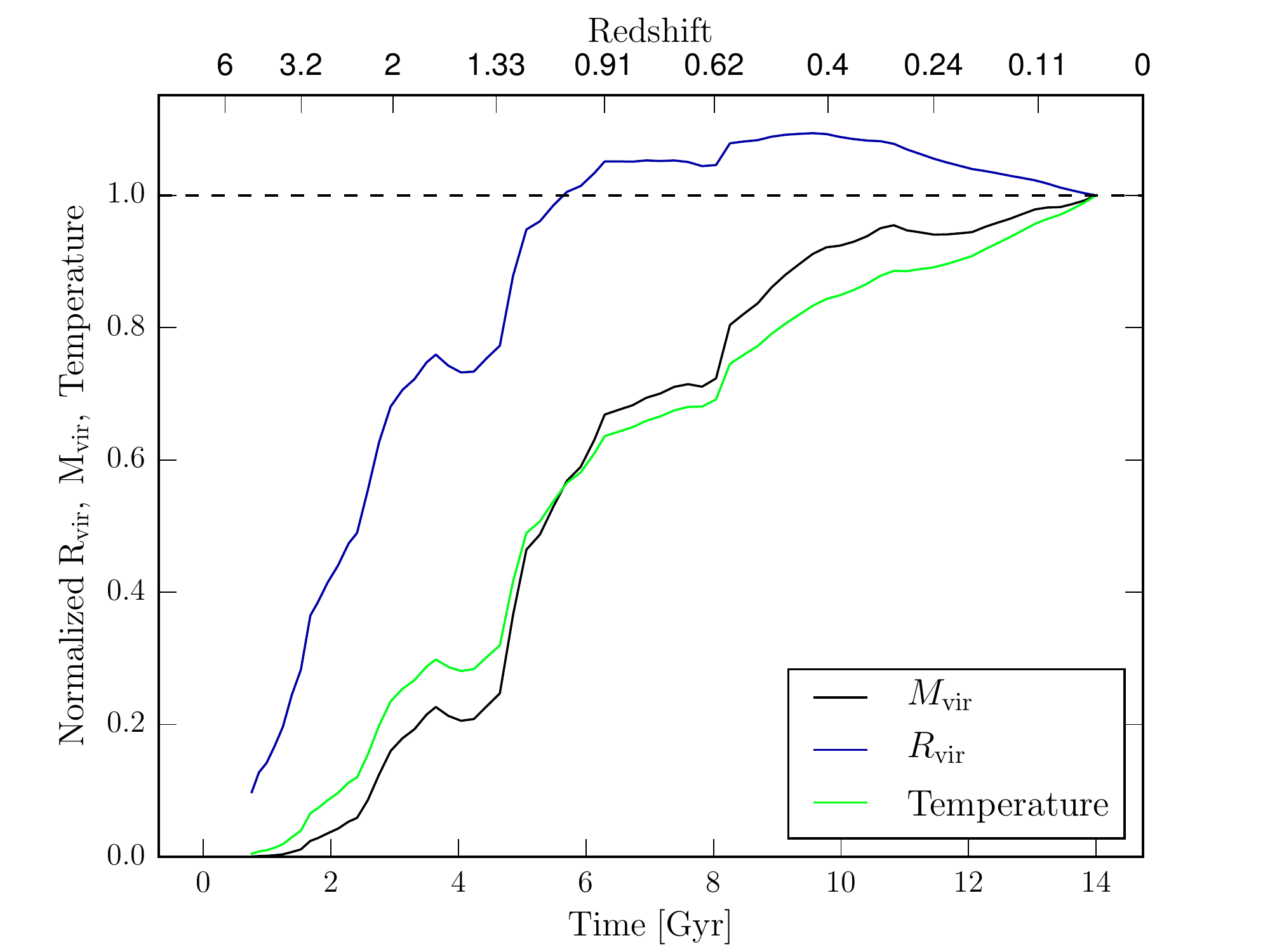}
       \caption{Time evolution of the Milky Way parameters taken from the Louise galaxy ($M_\textrm{vir} = 10^{12}$ M$_\odot$ and $R_\textrm{vir}=261.3$ kpc at $z=0$) in the ELVIS\citep{garrison-kimmel_elvis:_2014} simulations.
         The temperature is computed assuming a virial equilibrium at all time following equation \ref{eq:vir_temp}.
         \label{fig:evolving_milky_way}
       }
     \end{center}
   \end{figure}

   In the evolution mode \emrhoT, in addition to the density, we evolve the temperature $T$ as show in Fig.~\ref{fig:evolving_milky_way}.
   $T$ is computed assuming a virial equilibrium of the halo gas at any time, using the following equation:
   \begin{equation}\label{eq:vir_temp}
     T = \frac{2}{5} \frac{m_\textrm{p} G M_\textrm{vir}(t)}{R_\textrm{vir}(t) k_\textrm{b}},
   \end{equation}
   where $M_\textrm{vir}(t)$ and $R_\textrm{vir}(t)$ are respectively the time-evolving virial mass and radius, 
   $m_\textrm{p}$  the proton mass, $G$  the gravitational constant,  and $k_\textrm{b}$ is the Boltzmann constant.
   The Plummer softening parameter $a$ is chosen in order to match the scale radius $R_s = R_\textrm{vir}/c$.
   
   \subsection{Satellite orbits}\label{sec:sim_orbit}

   For the moving box simulations, we used only one generic orbit for the satellites galaxies.
   A deeper analysis of the influence of the orbital parameters on the dwarfs has been previously done with \texttt{GEAR} in \citet{nichols_gravitational_2014}.

   According to recent proper motions and orbital parameters determination of dwarf galaxies based on the Gaia DR2 \citep{fritz_gaia_2018}, confirming earlier studies 
   \citep{piatek_proper_2003,piatek_proper_2007},
   classical dwarfs such as Carina, Sextans and Sculptor have orbits with perigalacticon between $40$ and $120$~kpc and apogalacticon between $90$ and $270$~kpc \citep{fritz_gaia_2018}.
   It is worth noting that those measurements allow a fairly large interval of the orbital parameters.
   Therefore we decided to use a generic orbit with a pericentre of $60$~kpc and an apocentre of $150$~kpc, together with a current position of the dwarf at a distance of $85$~kpc, 
   with a negative velocity along the radial axis.

   We emphasize here that wind tunnel simulations are very complementary to the moving box approach.  
   Indeed they allow to explore a much larger parameter space of the hot gas temperature and density and 
   infalling velocity of the satellites, than could be efficiently done with the moving boxes. In that respect, 
   one does not need to sample a very large sets of orbits, as those would duplicate the parameters investigated by 
   the wind tunnels. 
   
\begin{figure}
     \begin{center}
       \includegraphics[width=0.5\textwidth]{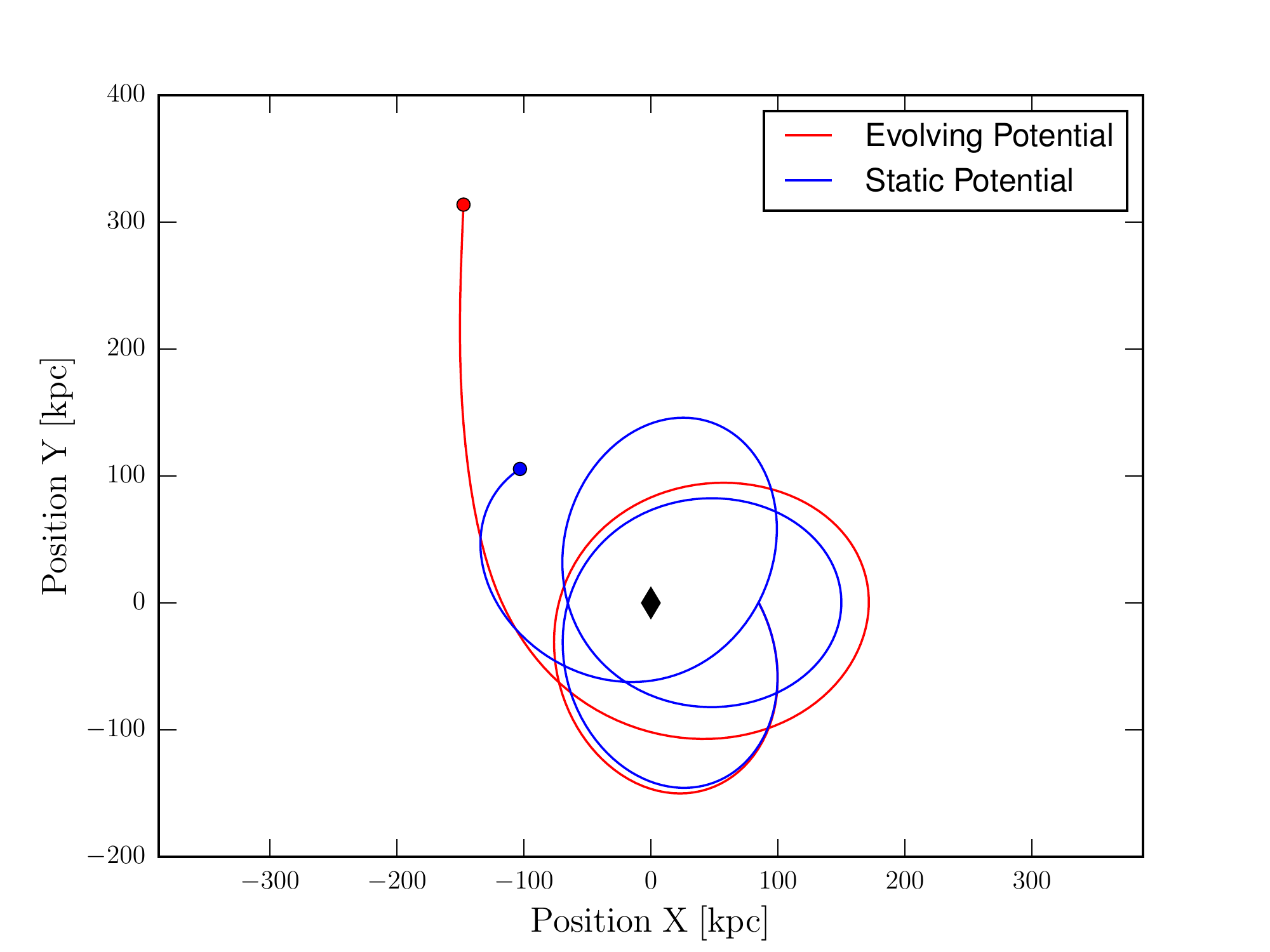}
       \caption{Orbit used for the static (\emsta~ mode) and evolving (\emrho~ and \emrhoT~ modes) MW potentials. The black diamond indicates the potential centre. The two points show the initial position at $z=z_{\rm{ext}}=2.4$.
         The final position is the same for both potentials and is situated at the coordinate $[0, 85]$~kpc.
       \label{fig:orbit}}
     \end{center}
   \end{figure}

   We also recall that due to the constraints imposed by the moving box method (See Section~\ref{sec:implementation}), we are unable to use orbits with a pericentre smaller than $30$~kpc, as
   the latter must be larger than half of our box size.
   To get the initial position of the extracted satellite at the infall time, $z=z_{\rm{ext}}$, the orbit of a point mass is backward time-integrated in both the static (\emsta~ mode) 
   and evolving (\emrho~ and \emrhoT~ modes) MW potential, using a Runge-Kutta algorithm.
   The two orbits obtained are compared in Fig.~\ref{fig:orbit}. In the static case, the satellite will perform two and a half orbit around the MW, while only one and a half in the evolving case.

   %------------------------------------------------------------%
   %------------------------- Section --------------------------%
   %------------------------------------------------------------%
   \section{Simulations}\label{sec:sim_set}
  
   \subsection{Wind tunnel simulations}
   Those simulations explore the effect of the wind parameters  on the evolution of the dwarf. Precisely, we explored its velocity relative to the dwarf $v_\textrm{w}$, 
   its temperature $T_\textrm{w}$ and density $\rho_\textrm{w}$. 
   We preformed in total 96 simulations corresponding to each combination of the wind parameters as presented in Tab.~\ref{table:wind_tunnel}.
   Each parameter is varied in a range of almost one dex around a fiducial value. They are set in order to match
   the observed MW constraints, either the gas  density \citep{miller_constraining_2015,miller_structure_2013} or the satellites velocities constraints by their proper motions
   \citep{piatek_proper_2003,piatek_proper_2007}.

   The fiducial parameters of the wind are chosen to match our static Milky Way model at injection position.
   They are set to a density $\rho_\textrm{w}=1.66\cdot 10^{-5}$ atom/cm$^3$, a velocity $v_\textrm{w}=100$ km/s and a temperature $T_\textrm{w}=2\cdot 10^6$K).
   The bottom line of Tab.~\ref{table:wind_tunnel} indicates the ratio of the parameters with respect to the fiducial ones. 
      
   As presented in Sec.~\ref{sec:dwarfs}, we exposed the dwarf model \texttt{h159} to the wind.  This galaxy presents a rather shallow gravity potential, therefore the RP is 
   efficient at stripping the gas and makes it sensitive to the wind parameters. 
   Its initial cold ($T\le1000\,\rm{K}$) and hot ($T<1000\,\rm{K}$) gas mass at infall time, $z=z_{\rm{ext}}$ is respectively $4.58$ and $27.0 \cdot 10^6\,\rm{M}_\odot$.
   6 more massive galaxies have been also simulated (see Tab.~\ref{tab:addwt}) confirming results obtained by model \texttt{h159}.

   \begin{table*}
     \begin{center}
       \caption{Wind parameters used in the wind tunnel simulations. 
       	The bottom line indicates the ratio of each parameter with respect to its corresponding fiducial one. The fiducial parameters are given in the fourth column.}
       \begin{tabular}{c|cccccc}
         \hline\hline
         $v_\textrm{w}$ [km/s]                     & -       & $76.9$    & -        & $100$   & $130$ & $169$ \\
         $T_\textrm{w}$ [$10^6$K]                  & $1.30$  & $1.54$    & $1.76$   & $2.0$\phantom   & $2.60$   & $3.39$ \\
         $\rho_\textrm{w}$ [$10^{-5}$ atom/cm$^3$] & -       & $1.28$    & -        & $1.66$  & $2.16$  & $2.81$ \\ \hline
         ratio to the fiducial parameter           & 0.65    & 0.77      & 0.88    & 1.      & 1.3    & 1.69 \\ \hline\hline
       \end{tabular}
       \label{table:wind_tunnel}
     \end{center}
   \end{table*}

   \subsection{Moving box simulations}
   Those simulations explore the impact of the MW on the evolution of dwarf galaxies through a most complete interaction model
   which takes into account the orbits of the dwarf satellite through a time-variation of the wind parameters, but also the gravitational tidal effects
   together with the mass growth of the MW over a Hubble time.
   
   We studied the evolution of 7 dwarfs, with total halo masses from $M_\textrm{200} = 5.4$ to $26.2 \cdot 10^8$ M$_\odot$. In the following, we will only focus on two 
   representative models, the quenched model \texttt{h159} dominated by old stellar populations and the Sculptor-like model \texttt{h070} which has an extended star formation history.
   Other models, including more massive ones characterized by a sustained star formation rates give similar results. See Tab.~\ref{tab:addmb} for the list of additional models simulated.
   In a first step, each of these two dwarfs have been simulated in isolation. In a second step, they have been simulated in the three modes including the Milky Way interaction, 
   \emsta, \emrho\, and \emrhoT.
   
   In Table \ref{table:real_sim}, the parameters of each moving box and isolated fiducial simulations are given.

   \begin{table}
     \begin{center}
       \caption{Description of the realistic simulations.
          The dwarf model names come from \citet{revaz_pushing_2018} supplemented by the MW model as described in \ref{sec:milky_way}.
          If no MW model are given, it means that the simulation was done in isolation and therefore do not contain a host.
         \texttt{h159} displays a quenched star formation history while the one of \texttt{h070} is extended.
         \label{table:real_sim}}
       \begin{tabular}{l|cl}
         \hline\hline
         Name                  & Dwarf Model   & MW Model          \\ \hline
         \texttt{h159\_iso}    & \texttt{h159} & -                 \\
         \texttt{h159\_sta}    & \texttt{h159} & \emsta \\
         \texttt{h159\_rho}    & \texttt{h159} & \emrho    \\
         \texttt{h159\_tem}    & \texttt{h159} & \emrhoT  \\ \hline
         \texttt{h070\_iso}    & \texttt{h070} & -                 \\
         \texttt{h070\_sta}    & \texttt{h070} & \emsta \\
         \texttt{h070\_rho}    & \texttt{h070} & \emrho    \\
         \texttt{h070\_tem}    & \texttt{h070} & \emrhoT   \\ \hline \hline
       \end{tabular}
     \end{center}
   \end{table}

   %------------------------------------------------------------%
   %------------------------- Section --------------------------%
   %------------------------------------------------------------%
   \section{Results}\label{sec:results}

   %------------------------------------------------------------%
   %------------------------- Subection --------------------------%
   \subsection{Analysis}\label{sec:analysis}

   \subsubsection{Pressure ratio}
   During the infall of a dwarf galaxy towards its host, the hot halo gas of the latter not only exerts a ram pressure against the ISM of the former, 
   but also an almost uniform thermal pressure (TP) all around it.
   A key point to understand how the dwarf galaxy evolution is impacted upon infall, is to measure the individual effect of both the RP and TP, as they 
   both have an opposite effect.
   While the RP removes the gas from the galaxy by momentum transfer,
   the TP tends to protect it by applying an additional force all around it, which prevents its removal due to RP, SNe feedback or UV-background heating
   resulting from the UV-photons emitted by active nuclei and star-forming galaxies.
   As presented by \citet{sarazin_x-ray_1986}, the ram pressure is given by
   \begin{equation*}
     P_\textrm{RP} = \rho_\textrm{w} v_\textrm{w}^2,
   \end{equation*}
   where $\rho_\textrm{w}$ is the wind density and $v_\textrm{w}$ its velocity.
   The thermal pressure is given by the ideal gas law
   \begin{equation*}
     P_\textrm{TP} = n k_\textrm{B}T_\textrm{w},
   \end{equation*}
   where $n$ is the particle number density, $k_\textrm{B}$ the Boltzmann constant and $T_\textrm{w}$ the wind temperature.
   Consequently, ratio of TP and RP which defines a unitless coefficient is written as
   \begin{equation}\label{eq:beta}
     \beta_\textrm{RP} = \frac{k_\textrm{B}}{\mu m_\textrm{P}}\frac{T_\textrm{w}}{v_\textrm{w}^2},
   \end{equation}
   where $\mu$ the mean molecular mass and $m_\textrm{P}$ the proton mass.

   We will see that this ratio will play a crucial role in the analysis and understanding of our simulations.
   It is worth noting that in Eq.~(\ref{eq:beta}), the density disappears and therefore the RP striping is independent of it at first order.

   \subsubsection{Gas fraction computation}\label{sec:gas_frac}
   In order to estimate the effect of RP stripping, we compute the gas fraction of our dwarf galaxies with time.
   It is performed by computing the mass of the hot gas in a constant radius taken as the initial virial radius $R_\textrm{200}(z_\textrm{init})$.
   As contrary to the hot gas, the cold gas is concentrated around the dwarf centre, we computed the cold gas mass in a radius $R_{cg}$ equal to 
   10\% of $R_\textrm{200}(z_\textrm{init})$.

   %------------------------------------------------------------%
   %------------------------- Subection --------------------------%

   \subsection{Wind tunnel simulations}\label{sec:windtunnel_results}
 
   Our wind tunnel simulations confirm the strong effect the hot halo gas has on the dwarf ISM through RP.
   However they also reveal the importance of the satellite's ISM multiphase structure. 
   In a first step, we therefore split our analysis according to the gas temperature.
   In a second step, we will explore the effect on the star formation and study the impact of the wind parameters.
   A short summary will be given at the end of the section.

   \subsubsection{Stripping of the hot gas}\label{sec:hot_gas_stripping}
\begin{figure*}
     \centering
     \subfloat{\includegraphics[width=0.24\textwidth]{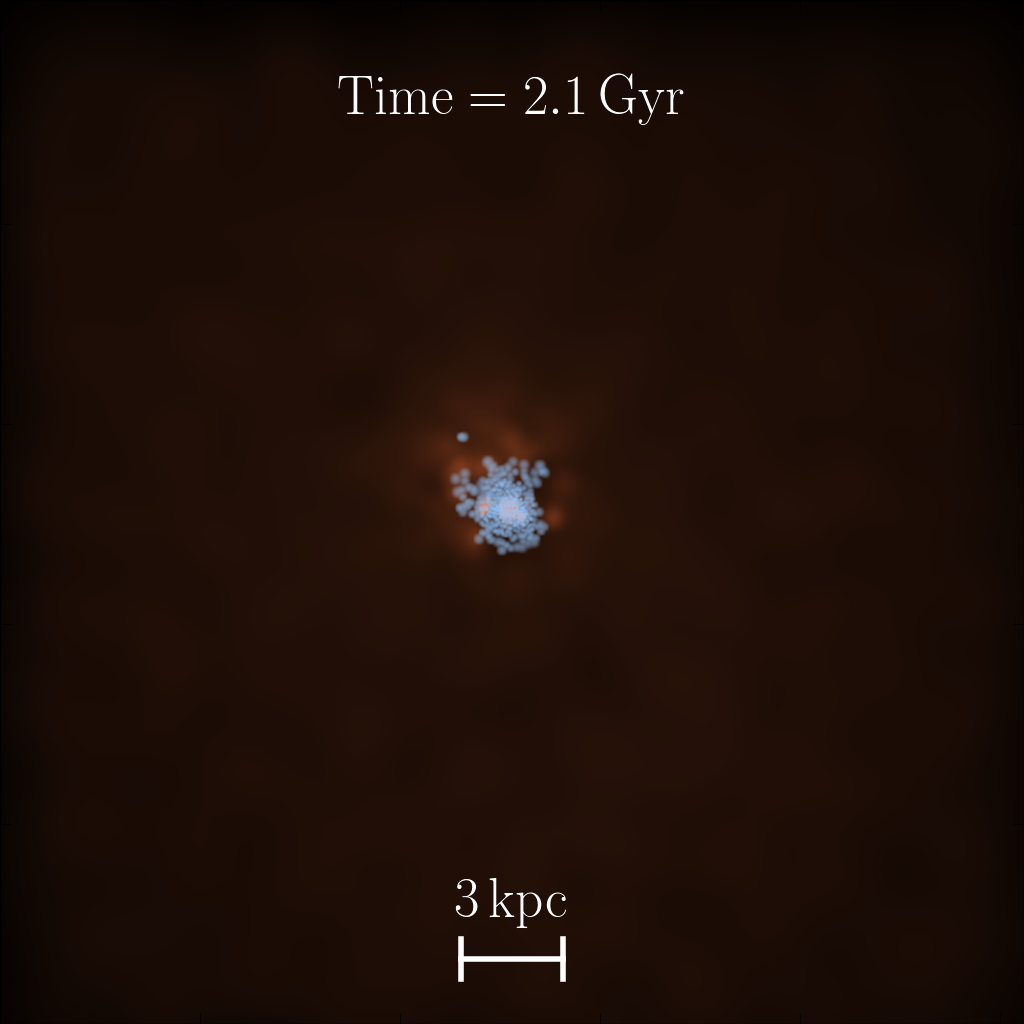}}
     \subfloat{\includegraphics[width=0.24\textwidth]{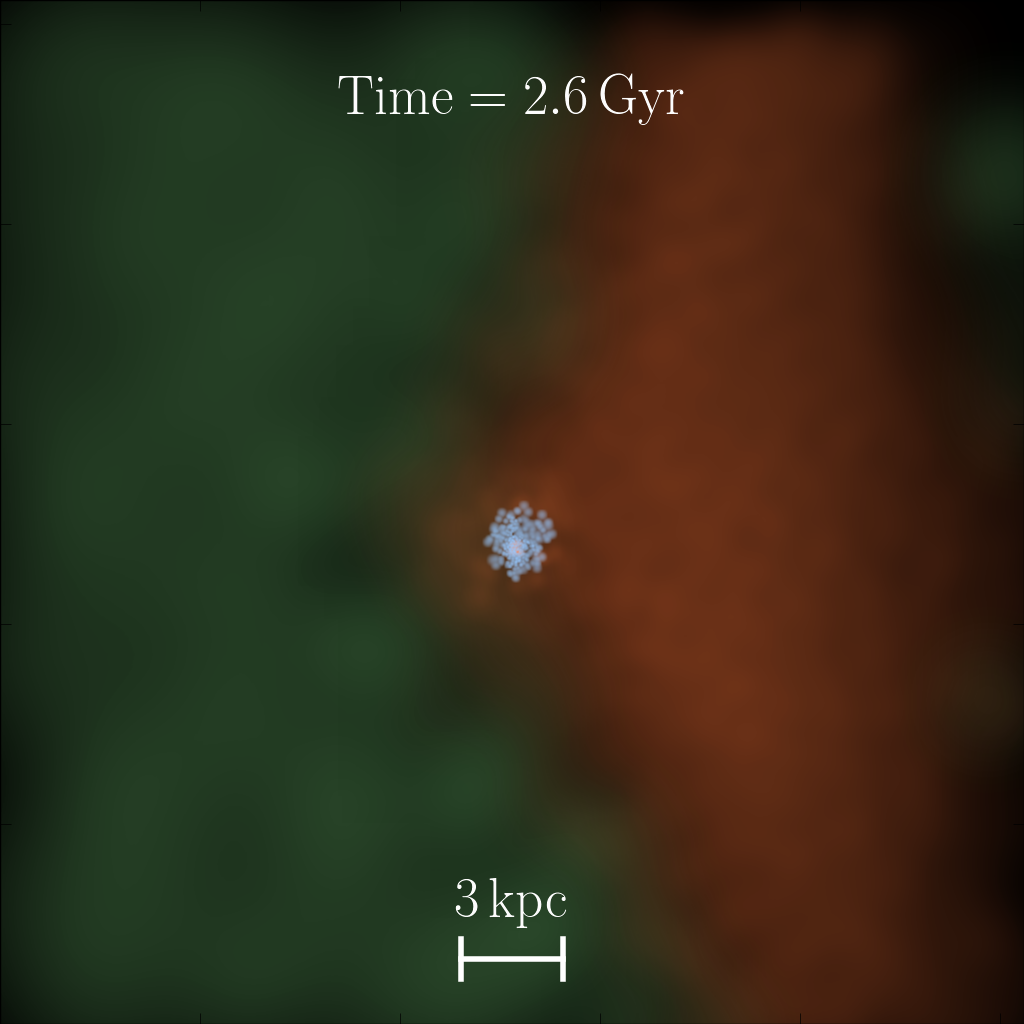}}
     \subfloat{\includegraphics[width=0.24\textwidth]{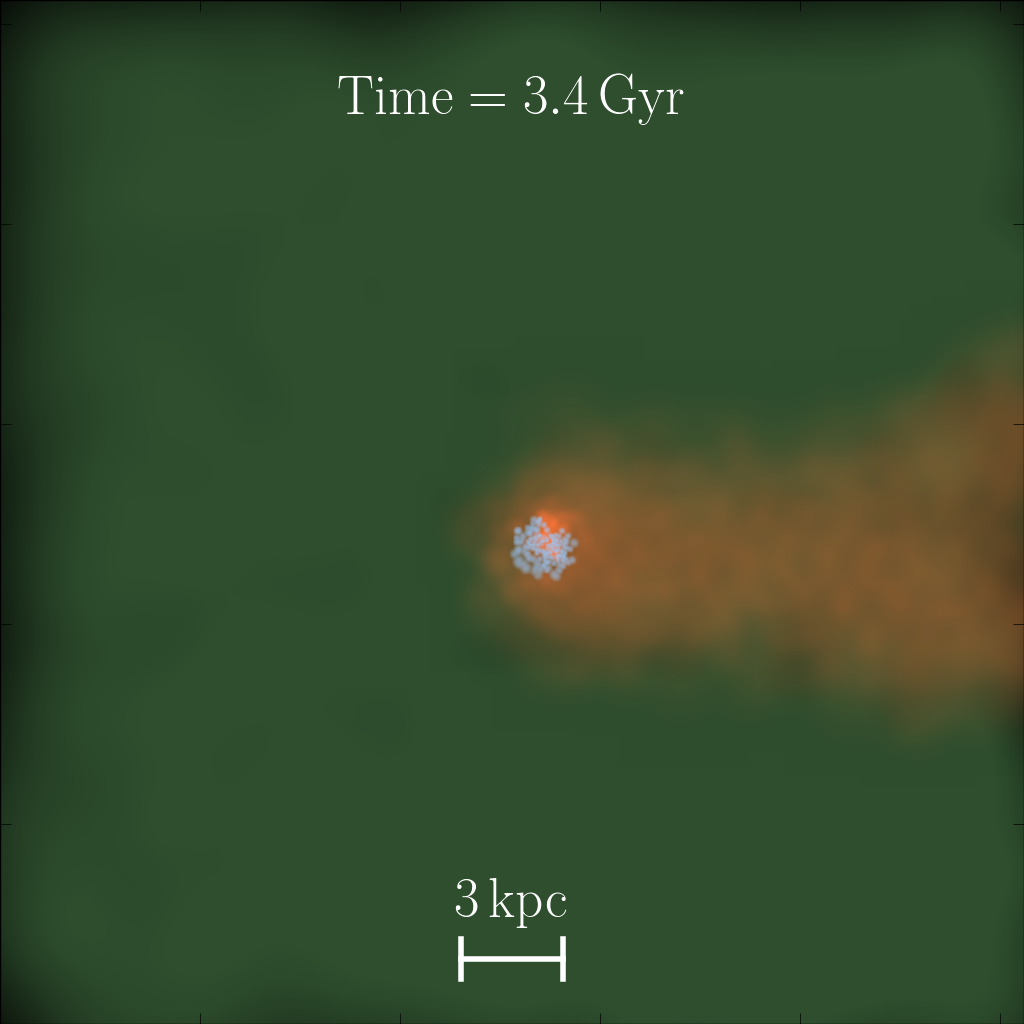}}
     \subfloat{\includegraphics[width=0.24\textwidth]{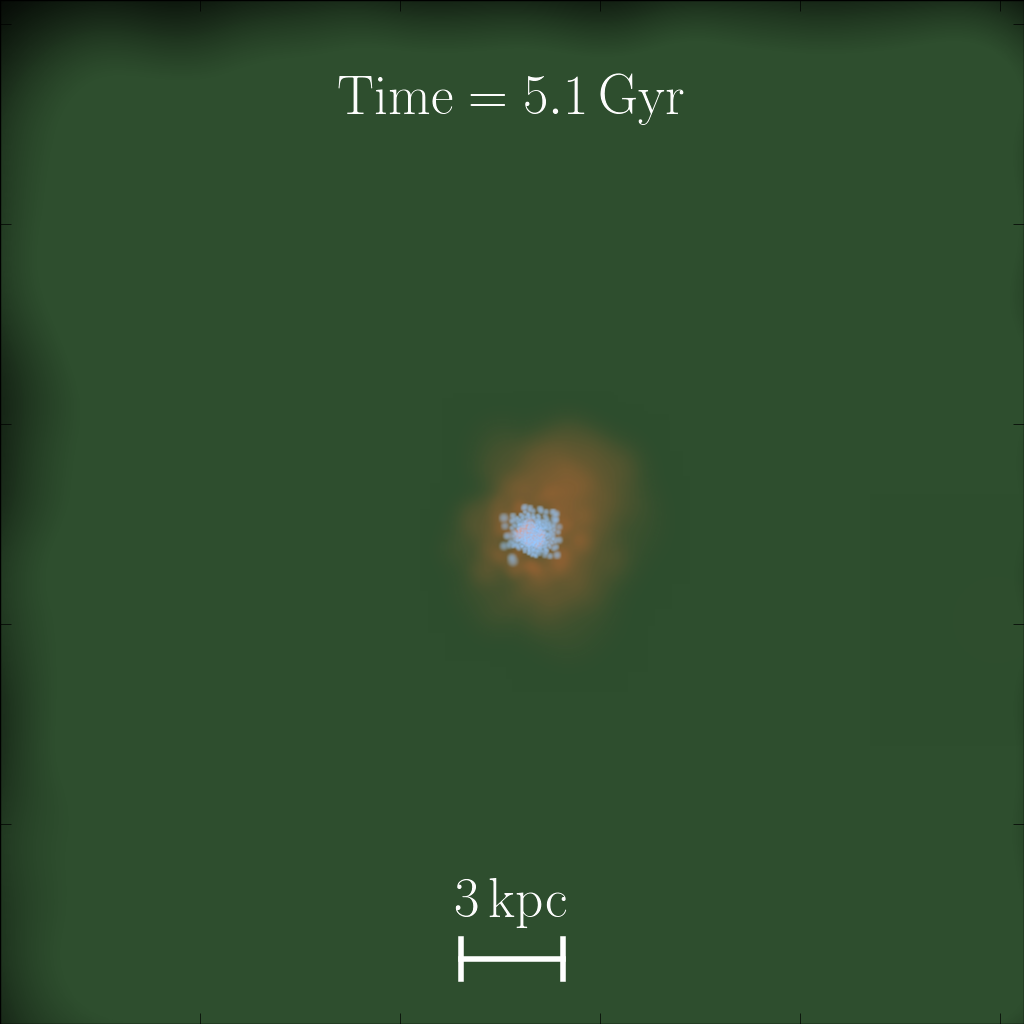}}
     \caption{Evolution of the cold, hot and wind gas during the first contact between the dwarf galaxy and the hot halo in a wind tunnel simulation with a wind temperature of $3.39$~K, a density of $1.28 \cdot 10^{-5}$~atom/cm$^3$ and a velocity of $76.9$~km/s.
       The hot gas of the dwarf ($T_\textrm{w} > 10^3\,\rm{K}$) is shown in red, its cold gas ($T_\textrm{w} < 10^3\,\rm{K}$) in blue. The green colours trace the gas of the wind.
       The green and black lines correspond to the isolated and fiducial wind tunnel model respectively.
       The blue dashed lines indicate the galaxy injection time.
       This sequence shows how the hot dwarf gas is quickly stripped while its cold gas remains.
     \label{fig:rps_evolution}}
   \end{figure*}

\begin{figure*}
	
	\centering
	\leavevmode   	
	\subfloat[hot gas fraction]{\resizebox{0.49\hsize}{!}{\includegraphics[angle=0]{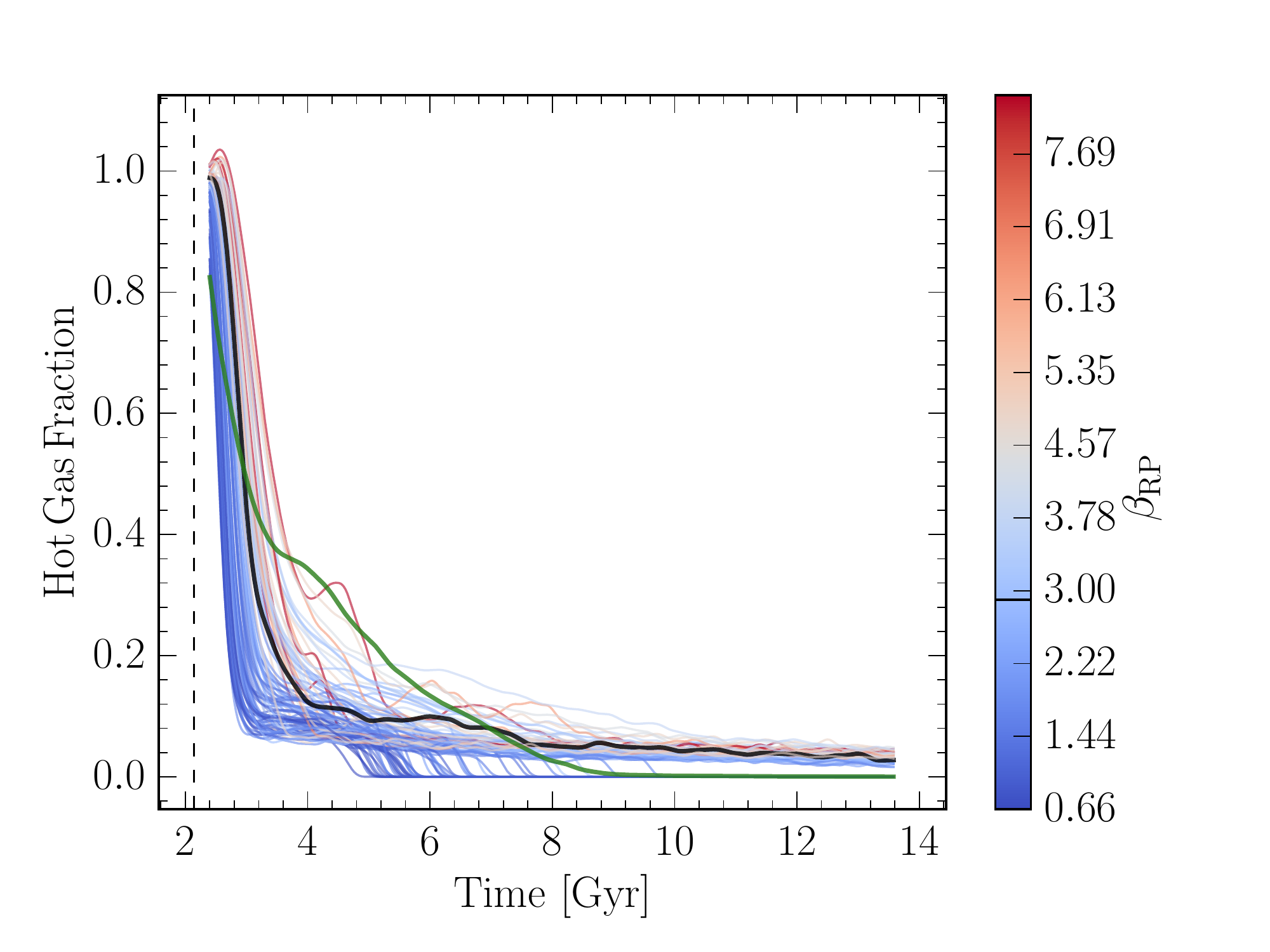}}}
	\subfloat[cold gas fraction]{\resizebox{0.49\hsize}{!}{\includegraphics[angle=0]{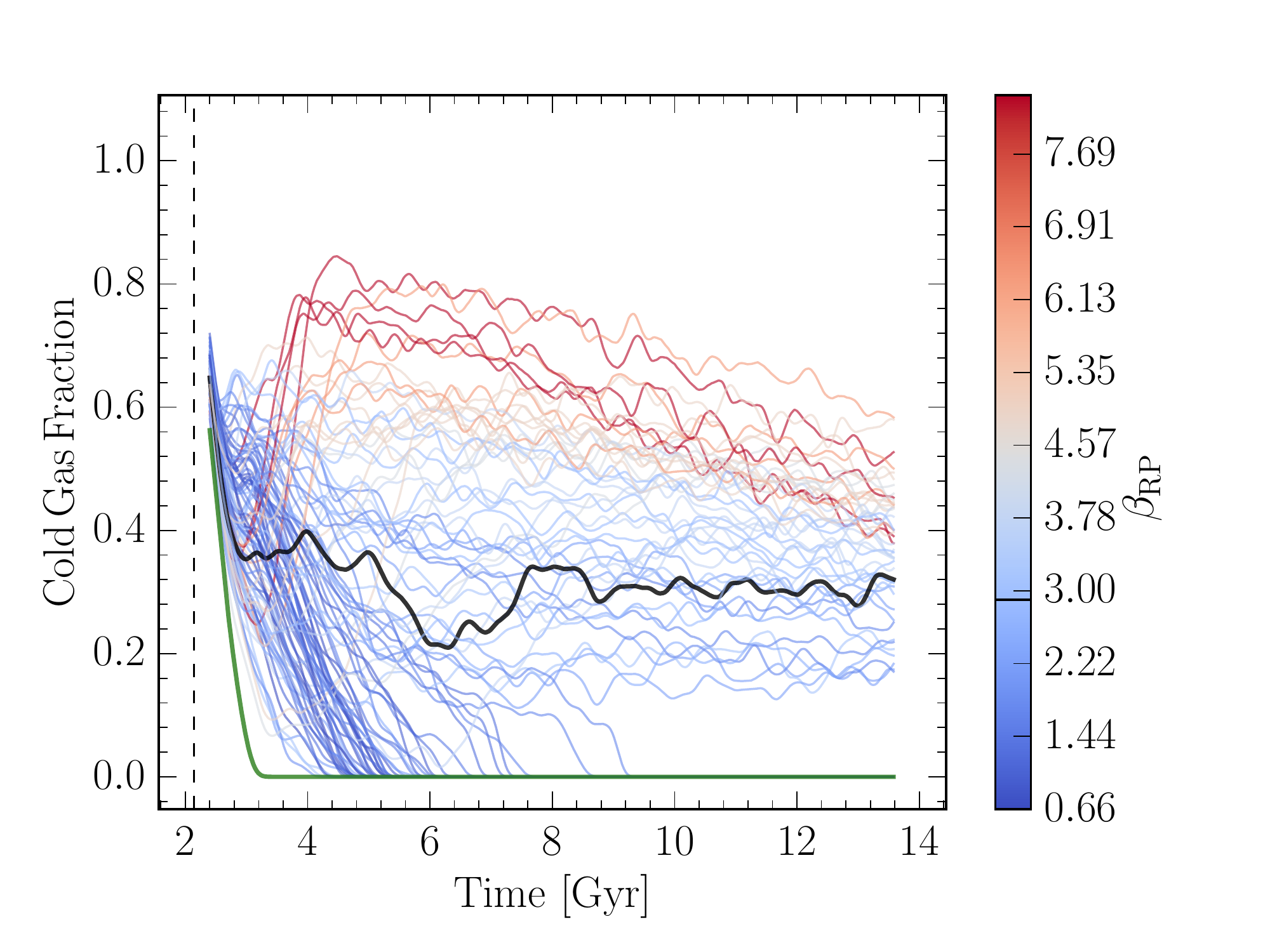}}}

        \caption{
          Time evolution of the gas fraction of the dwarf galaxy \texttt{h159} evolving through a wind tunnel simulation (see Table \ref{table:wind_tunnel} for the list of parameters).
          left panel : the hot gas fraction contained in one virial radius.
          right panel : the cold gas fraction contained in 0.1 virial radius.
          The colour of each line reflect the corresponding $\beta_{\rm{t}}$.
          In both panels, the green and black lines correspond to the isolated and fiducial wind tunnel model respectively.
          A moving average has been applied with a gaussian kernel (standard deviation of $\sim 100\,\rm{Myr}$ in a window of $-500$ to $500\,\rm{Myr}$) to reduce the noise.
          Due to this filter, the earliest times are removed and the different curves start at different fraction.
        }
        \label{fig:f_gas}
	
\end{figure*}

   Figure~\ref{fig:rps_evolution} shows the evolution of model \texttt{h159} exposed to a wind of temperature equal to $3.39 \cdot 10^6$~K, a density of $1.28 \cdot 10^{-5}$~atom/cm$^3$ and a velocity of $76.9$~km/s.
   This time sequence shows four different important steps.
   The first frame shows the gas at $t=2.1$ Gyr, before any hydrodynamic interaction between the wind and the dwarf. 
   The second one shows the first contact,
   the third one shows the state of the dwarf about one Gyr after the first contact. The last one corresponds to the steady state reached after the RP stripping.   
   As expected, soon after the first contact, the large hot halo gas of the dwarf is strongly distorted ($t=2.6\,\rm{Gyr}$) and quickly stripped, forming a trailing
   tail beyond the dwarf ($t=3.4\,\rm{Gyr}$). At later time, only a small hot halo gas remains around the dwarf. The latter was not initially part of the dwarf halo gas.
   It results from the permanent heating of the cold gas by both UV-background heating and supernovae feedback.
   The efficient stripping of the hot gas is confirmed by the left panel of Fig.~\ref{fig:f_gas} where the time evolution of the hot gas fraction is shown for all of
   our 96 wind tunnel simulations. The colour of each line corresponds to the parameter $\beta_\textrm{RP}$, the ratio between the thermal and ram pressure (Eq.~(\ref{eq:beta})).
   All simulations show a quick drop of their hot gas fraction, indicating the efficient stripping of the dwarf hot halo. This demonstrates that the ram pressure stripping
   is captured in our simulations. The left panel of Fig.~\ref{fig:f_gas} also reveals a weak dependency on $\beta_\textrm{RP}$. Winds characterized by a smaller $\beta_\textrm{RP}$ 
   are more efficient to ram pressure strip the hot dwarf gas. 
   Finally, we see that the isolated case traced by the green curve retains more hot gas after
   $4\,\rm{Gyr}$ as the latter do not suffer any ram pressure stripping.
   However, at later time the warm gas fraction decreases. This reveals the secular 
   evaporation of the hot gas due to the continuous UV-background heating, until complete evaporation at $t\cong 9\,\rm{Gyr}$.
   The remaining of hot gas in the wind tunnel simulations after that time compared to the isolated model will be discussed below.

   \subsubsection{Stripping of the cold gas}
   Contrary to the hot dwarf gas, the cold one is much more difficult to strip. This is well observed on the last panel of Fig.~\ref{fig:rps_evolution} at $t=5.1\,\rm{Gyr}$, where even $3\,\rm{Gyr}$ after the first contact, cold gas is still present in the dwarf.
   The right panel of Fig.~\ref{fig:f_gas} shows in more detail, the time-evolution of the cold gas fraction for all our wind tunnel simulations. 
   We split our models in two categories according to their $\beta_\textrm{RP}$ value:
   (i) thermal pressure-dominated models : $\beta_\textrm{RP} \ge \beta_{\rm{t}}$, red colours,
   (ii) ram pressure-dominated models : $\beta_\textrm{RP} < \beta_{\rm{t}}$, blue colours, 
   where $\beta_{\rm{t}}$ is defined as the value at the transition and is about $3$ for this galaxy.

   Understanding the evolution in these different regimes first requires comprehension of the cold gas evolution in the isolated case.
   On the right panel of Fig.~\ref{fig:f_gas} the corresponding cold gas fraction is traced by the green curve. It is striking to see that the latter is 
   dropping quickly, in less than $4\,\rm{Gyr}$, faster than any other wind tunnel model. 
   As for the hot gas, the origin of this drop is due to the UV-background ionizing photons which heat the gas. The potential well
   of this dwarf model being shallow, the latter evaporates \citep{efstathiou_suppressing_1992,quinn_photoionization_1996,bullock_reionization_2000,noh_physical_2014} resulting in the
   star formation quenching of the galaxy \citep{revaz_pushing_2018}.

   \myparagraph{Thermal pressure-dominated models ($\beta_\textrm{RP} > \beta_{\rm{t}}$)}
   When the thermal pressure dominates over the ram pressure, the high pressurized wind compress the cold gas, protect it against 
   ram pressure and act against its UV-background heating driven evaporation observed in the isolated case.
   This protection leads to keep up to 50\% of cold gas, even after a Hubble time.
   The regular decrease of the mass fraction observed  in this regime
   results from the conversion of the cold gas in to stars resulting from a continuous star formation rate.
   This point will be discussed further below.   
   
   Low wind velocity models show an important drop of the cold gas fraction followed by a strong rise between $2$ and 
   $4\,\rm{Gyr}$. The drop results from some gas particles being pushed by the wind, leaving the cut off 
   radius, where the cold gas is measured. However, those particles do not acquire enough kinetic energy to leave the 
   galaxy and are thus slowly re-accreted by gravity, explaining the subsequent increase of the mass fraction.

   The oscillations observed in nearly all models result from the continuously pulsation of the ISM induced by the numerous supernovae explosion
   which cause the gas to be ejected outwards $R_{cg}$ (the radius used to compute the cold gas) before being slowly re-accreated.

   \myparagraph{Ram pressure-dominated models ($\beta_\textrm{RP} < \beta_{\rm{t}}$)}
   When the ram pressure dominates over the thermal pressure, the pressure protection is much weaker and the cold gas evolution becomes similar
   to the one of the isolated case.    
   While just below the transition $\beta_{\rm{t}}$, cold gas may still survive up to $z=0$,
   for very low $\beta_\textrm{RP}$, it is lost. 
   Those cases correspond to a fast moving dwarf with a speed larger than $150\,\rm{km/s}$
   entering the halo of its host with a temperature of at most $1.3\times 10^6\,\rm{K}$.
   However, in any case when the ram pressure is present, the cold gas fraction remains larger than the isolated case. This indicates that
   the UV-background heating always dominates over the RP.
   The final loss is due to a supernovae that ejects almost all the gas further than the stripping radius which is then removed from the galaxy as a single cloud.

   \subsubsection{Impact on star formation}\label{sec:sfr_impact}
   
\begin{figure}
     \begin{center}
       \includegraphics[width=0.5\textwidth]{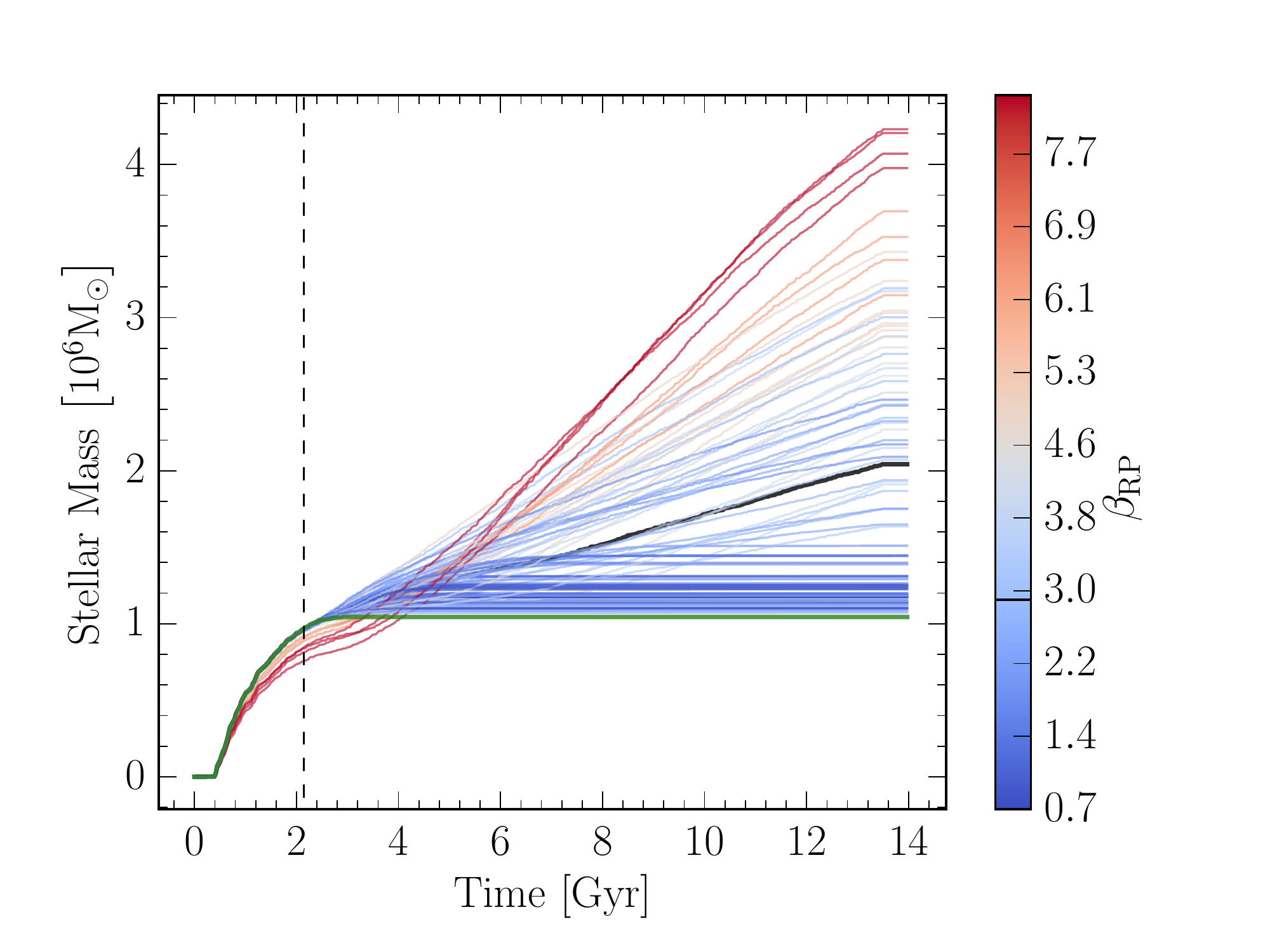}
       \caption{Stellar mass as a function of time for the wind tunnel simulations.
         The colour is defined by the coefficient $\beta_\textrm{RP}$ in equation \ref{eq:beta}.
         The black line corresponds to our fiducial wind parameters.
         The green line corresponds to the isolated case.
         The blue dashed line represents the injection time.
         \label{fig:sfr_beta}
       }
     \end{center}
   \end{figure}

   Together with an important change of the cold gas mass fraction with respect to the isolated model, our
   wind tunnel simulations strongly impact the star formation rate and subsequently the amount of stars formed.
   Fig.~\ref{fig:sfr_beta} displays the cumulative number of stars formed with time. Compared to the star formation
   history, this plot has the advantage of being much less noisy. 

   All wind tunnel 
   models form stars more efficiently compared to the isolated case as a consequence of the remaining large reservoir of cold gas.   
   For the extreme thermal pressure-dominated models, 
   the final stellar mass is up to four times larger than the isolated galaxy model while the ram pressure-dominated models with very low 
   $\beta_\textrm{RP}$ remains similar.
   It is worth nothing that pressure-dominated models with very low $\beta_\textrm{RP}$, the ones that lost all their cold gas before $z=0$,
   still display truncated star formation histories, however, much more extended than the isolated case, up to $9\,\rm{Gyr}$ in the most extreme case
   
   For the sake of clarity, we note a small difference between all models, in the amount of stars formed before the injection time at 
   $z=z_{\rm{ext}}$, indicated by a vertical dashed line. Indeed, in order to computed the evolution of the stellar 
   mass, we extracted all stellar particles in the dwarf at $z=0$ and used their age to deduce the stellar mass present 
   at any comic time. Consequently, star particles formed in the dwarf but leaving the galaxy at later time are no 
   longer uncounted for, which may induce a small bias and the scatter observed between the different models. 
   This approach, contrary to others where the stellar mass is computed at any time during the evolution, 
   is much more representative to what an observer would have obtained relying on stellar ages deduced from a colour-magnitude diagram
   at present time.

   \subsubsection{Effect of the wind parameters}
      
\begin{figure*}
     \begin{center}
       \includegraphics[width=0.9\textwidth]{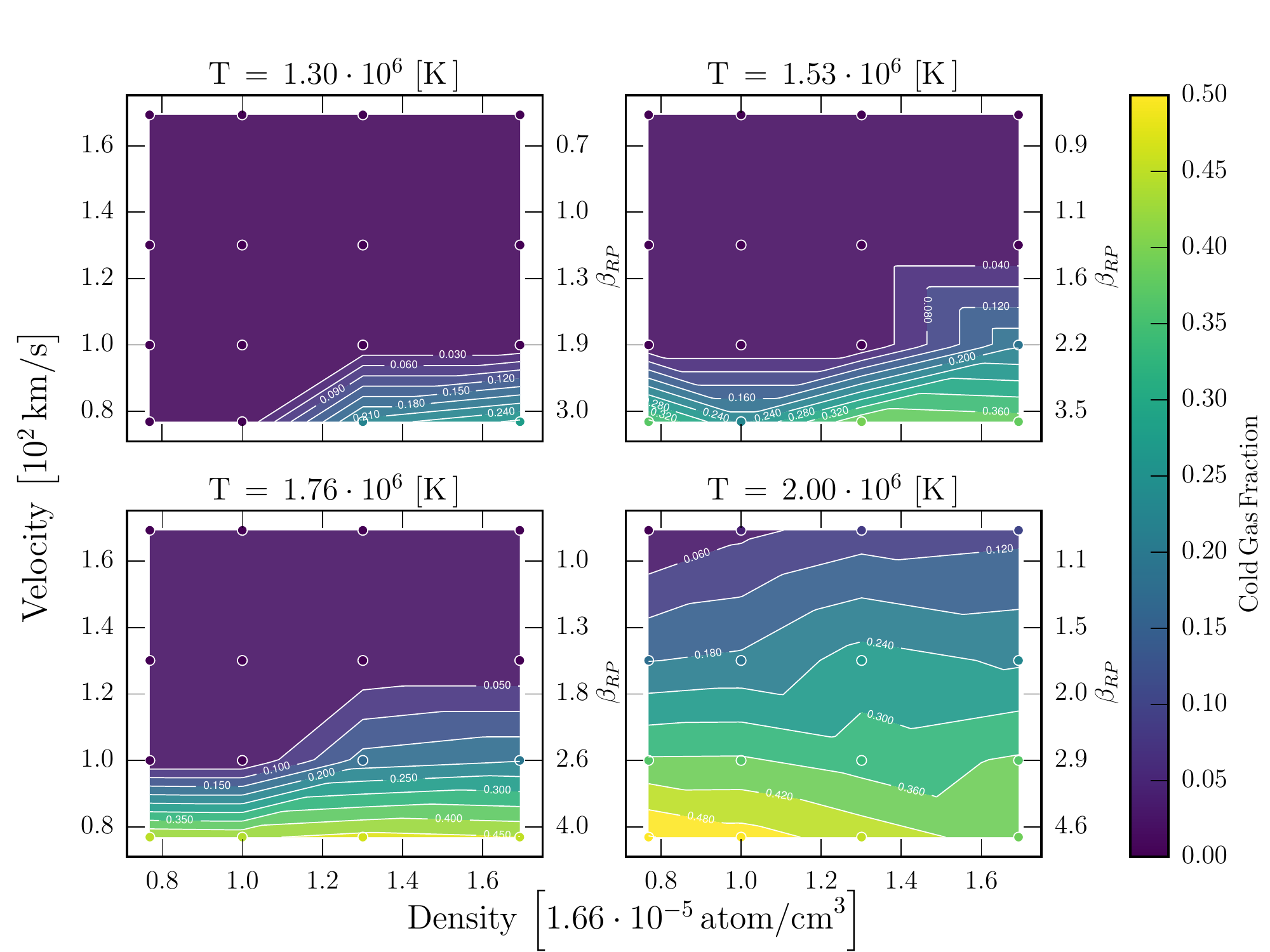}
       \caption{Each plot represents the cold gas fraction at $z=0$ of \texttt{h159} as a function of the hot halo density and satellite velocity for different halo temperatures.
         The white circles are the simulations done.
         $\beta_\textrm{RP}$ is given on the right axis of each graph.
         Usually, the RPS is described only through density and velocity, but here the temperature dependency is shown to have an important impact.
         \label{fig:dep}}
     \end{center}
   \end{figure*}

   Figure~\ref{fig:dep} shows the final cold gas fraction of our 96 wind tunnel simulations, as a function of the wind parameters, more precisely, 
   its velocity ($v_\textrm{w}$), density ($\rho_\textrm{w}$) and temperature ($T_\textrm{w}$).
   According to Eq.~(\ref{eq:beta}), for a fixed temperature, the parameter $\beta_\textrm{RP}$ only depends on $v_\textrm{w}$, to the inverse of its square.
   We thus supplement the velocity-axis ($y$-axis) with its corresponding $\beta_\textrm{RP}$-value on the right of each plot.
   
   As expected in the theoretical formulas, $v_\textrm{w}$ and $T_\textrm{w}$ are the two most sensitive parameters.
   For any temperature bin, increasing $v_\textrm{w}$ from $80$ to $160\,\rm{km/s}$ move from a regime where the gas is protected, ending with
   an important cold gas mass fraction (between 0.3 to 0.5\%) to a regime where all the gas is evaporated and the galaxy is quenched.
   Similarly, increasing $T_\textrm{w}$ increases the thermal pressure which protect the dwarf gas.  
   In strongly thermal pressure-dominated regimes ($\beta_\textrm{RP} > \beta_{\rm{t}}$), a dwarf galaxy is thus able to protect its gas reservoir 
   from stripping, up to 50\%.
   On the contrary, from Eq.~(\ref{eq:beta}), the wind density has a limited impact on the galaxy gas fraction. However, it has a threshold effect.
   Indeed, for temperature between $1.3$ to $1.76\times 10^6\,\rm{K}$, below a density of about $1.2$ to $1.4\times 10^{-5}\,\rm{atom/cm^3}$, the ram pressure
   stripping is enhanced, for a fixed $v_\textrm{w}$ and $T_\textrm{w}$.   
   Extrapolating Fig.~\ref{fig:dep} to lower temperature, we can predict that a quenched dwarf, like our \texttt{h159} model, orbiting in a halo with a temperature 
   $T_\textrm{w} < 1.3\cdot 10^6\,\textrm{K}$ and a density $\rho_\textrm{w} < 2\cdot 10^{-5}$ atom/cm$^3$ will loose all its gas.
   
   \subsection{Moving box simulations}\label{sec:movingbox_results}
   In this section, we go one step further by supplementing our wind tunnel simulations with 
   tidal stripping induced by a realistic Milky Way model environment. We also study the time-variation of the wind parameters
   all along the dwarf orbit which reflects the inhomogeneous hot halo of the Milky Way but also its growth with time.   
   A summary of the final properties of the six simulations performed are given in Tab.~\ref{table:resultat_real_sim}.

      \begin{table*}
     \begin{center}
       \caption{Properties at $z=0$ of the dwarf models evolved in the moving box simulations.
         The model names fit the one of the corresponding dwarf in \citet{revaz_pushing_2018}.
         $R_{200}$ and $M_{200}$  corresponds to the virial radius and mass respectively. $L_\textrm{V}$ is the final V-band luminosity.
         The cold gas is defined as the gas with a temperature lower than $1000\,\rm{K}$ while the hot one with a temperature above.
         \label{table:resultat_real_sim}}
       \begin{tabular}{l|cccccc}
         \hline\hline
         Name               & $R_{200}$ [kpc] & $M_{200}$ [$10^8$ M$_\odot$] & $M_\star$ [$10^6$ M$_\odot$] & $L_\textrm{V}$ [$10^6$ L$_\odot$] & Cold Gas [$10^6$ M$_\odot$] & Hot Gas [$10^6$ M$_\odot$]\\ \hline
         \texttt{h159\_iso} & $19.4$          & $5.37$                       & $1.08$                       & $0.43$                            &  $0$                        & $9.05$ \\
         \texttt{h159\_sta} & $11.3$          & $1.06$                       & $2.47$                       & $1.04$                            &  $0$                        & $0.$ \\
         \texttt{h159\_rho} & $15.2$          & $2.56$                       & $2.61$                       & $1.78$                            &  $1.34$                     & $1.00$ \\
         \texttt{h159\_tem} & $14.1$          & $2.05$                       & $1.00$                       & $0.40$                            &  $0$                        & $0.$   \\ \hline
         \texttt{h070\_iso} & $26.3$          & $13.3$                       & $5.72$                       & $2.04$                            &  $0$                        & $20.6$ \\
         \texttt{h070\_sta} & $14.5$          & $2.22$                       & $18.5$                       & $10.9$                            &  $6.32$                     & $4.61$ \\
         \texttt{h070\_rho} & $20.3$          & $6.09$                       & $23.8$                       & $14.9$                            &  $6.23$                     & $6.93$ \\
         \texttt{h070\_tem} & $17.3$          & $3.76$                       & $5.42$                       & $1.92$                            &  $0$                        & $0.$   \\ \hline \hline
       \end{tabular}
     \end{center}
   \end{table*}

   \subsubsection{Effect of the Milky Way model}\label{sec:effect_of_mw_model}
   
\begin{figure*}
   	
       \centering
       \leavevmode   	
   	   \subfloat[\texttt{h159}]{\resizebox{0.49\hsize}{!}{\includegraphics[angle=0]{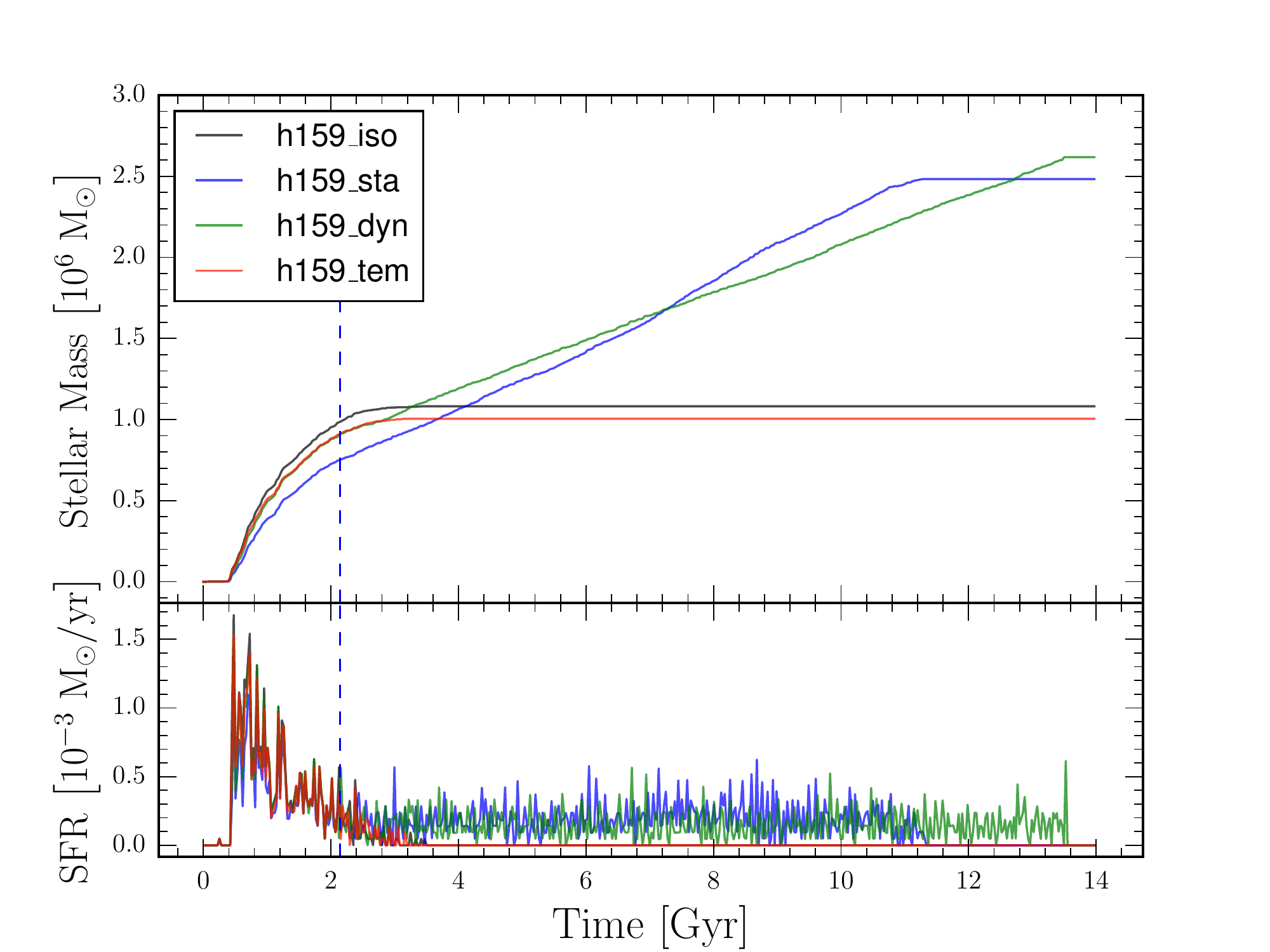}}}
   	   \subfloat[\texttt{h070}]{\resizebox{0.49\hsize}{!}{\includegraphics[angle=0]{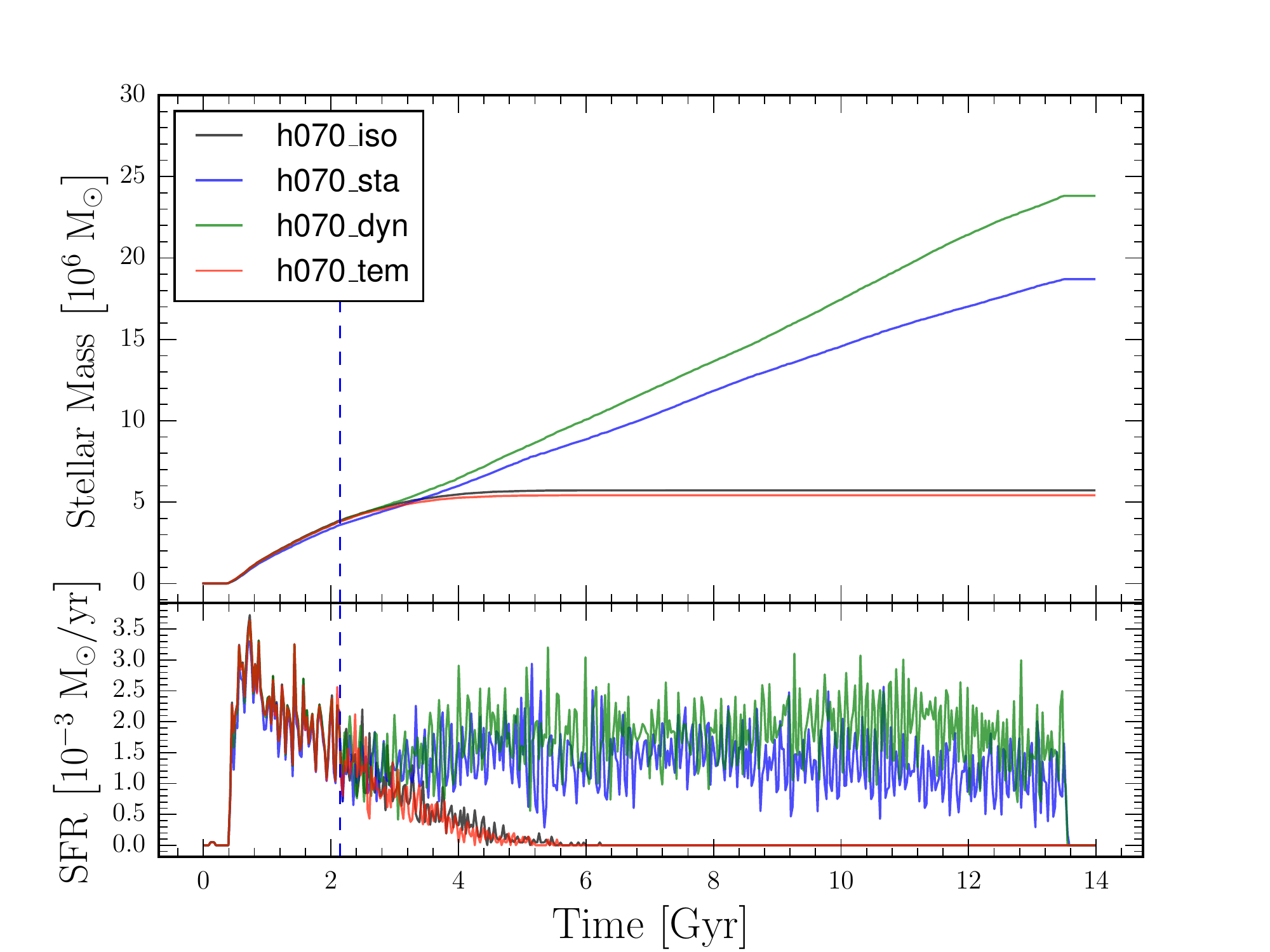}}}

       \caption{Time evolution of the cumulative stellar mass (top) and star formation rate (bottom) 
	   of model \texttt{h159} (left) and \texttt{h070} (right), in the four models: isolated (black), \emsta~ (blue),
	   \emrho~ (green) and \emrhoT~ (red). The vertical dashed line indicates the injection time for models \emsta, \emrho~ and \emrhoT.
	   %./pSFRvsTime.py -o BOTH --plot_time -1 --legend_loc "upper left" --xmin 0 --xmax 14 h159_iso h159_stat h159_dyn h159_dyn_T
	   %./pSFRvsTime.py -o BOTH --plot_time -1 --legend_loc "upper left" --xmin 0 --xmax 14 h159_iso h159_stat h159_dyn h159_dyn_T
       }   	
       \label{fig:sfr}
   
   \end{figure*}

\begin{figure}
     \begin{center}
       \includegraphics[width=0.5\textwidth]{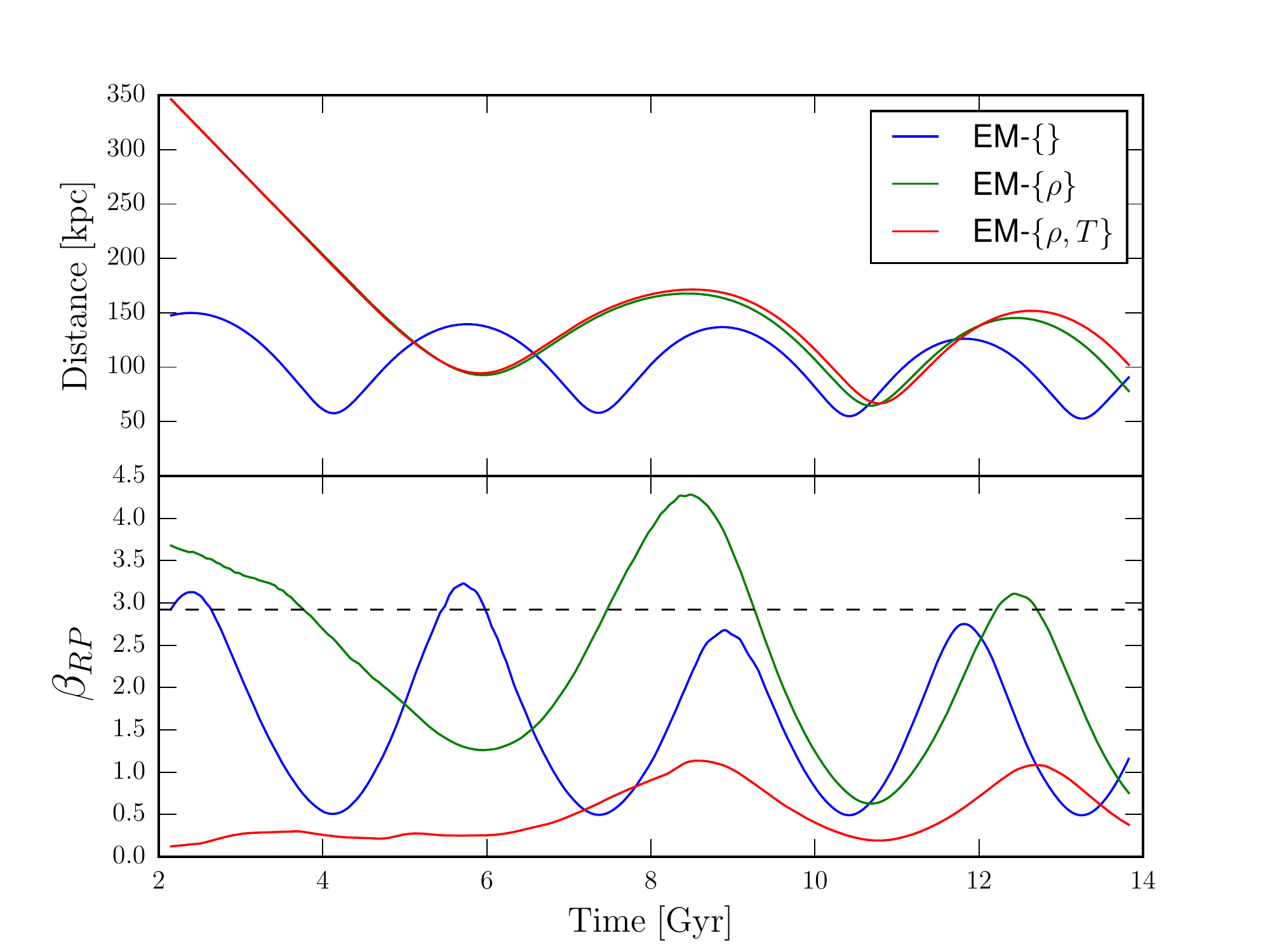}
       \caption{top panel: Time evolution of the distance of the dwarfs for both the models \texttt{h159} and \texttt{h070} with respect to their host galaxy centre.
       	bottom panel: The corresponding evolution of the $\beta_{\rm{RP}}$ parameter all along the dwarf orbit.
       	The blue, green and red curves correspond respectively to the \emsta, \emrho~ and \emrhoT.
        The black line represents our fiducial wind tunnel simulation (black line in Figure \ref{fig:f_gas}).
        % /home/epfl/lhausamm/archive/paper2018/plot_article/sfr_mw_beta/compute_beta.py
       	\label{fig:beta}}
     \end{center}
   \end{figure}

   Figure~\ref{fig:sfr} displays the star formation rate and time evolution of the cumulative stellar mass 
   for each of the four cases studied for the two dwarf models \texttt{h159} and \texttt{h070}, namely, isolated,
   \emsta, \emrho~ and \emrhoT.
   For the three last cases that include the ram pressure stripping, the bottom panel of Fig.~\ref{fig:beta} shows the corresponding evolution of the coefficient $\beta_\textrm{RP}$,
   while the top panel shows the distance of the dwarf with respect to its host galaxy (top panel).

   As discussed in \citet{revaz_pushing_2018}, when evolved in isolation, both models exhibit a star formation quenched 
   after respectively $\sim 3$ and $\sim 6\,\rm{Gyr}$ (black curves).   
   However, when the dwarfs enter a static Milky Way halo (\emsta, blue curve) at $z=z_\textrm{ext}$, the 
   star formation is no longer quenched but becomes continuous. As a consequence, the resulting final stellar mass is up to four times the one of the isolated case. 
   This increase of the star formation is due to the high $\beta_\textrm{RP}$ which pressurize the gas of the dwarf.
   As seen in Fig.~\ref{fig:beta}, $\beta_\textrm{RP}$ oscillates between 3 and 0.5 reflecting the dwarf orbit.
   Maximal values of 3, similar to our fiducial wind tunnel simulation, are reached during the apocentre passage, when the dwarf has the lowest velocity.  
   On the contrary, at the pericentre passage, at a distance of $50\,\rm{kpc}$, higher velocities increase the ram pressure with respect to the thermal one 
   and $\beta_\textrm{RP}$ drop down to 0.5.
   It is important to notice that even after four passages at the pericentre, the tidal force has not being strong enough to destroy the dwarf. This point is illustrated
   by the dark matter and stellar density profiles further discussed in Fig.~\ref{fig:all3}.
   
   The green curve (\emrho) corresponds to the case where the Milky Way increases its mass and density but keep a constant hot gas temperature.
   The Milky Way mass growth directly impacts on the dwarf orbit which experiments only three passages at the pericentre. It also impact on the $\beta_\textrm{RP}$
   parameter which starts with slightly higher values reflecting an initial larger distance ($\sim350\,\rm{kpc}$) and lower orbital velocity.
   As the density is initially much lower, a factor of about 30 compared to \emsta, the dwarf cold gas 
   is slightly less confined by the hot Milky Way halo (density threshold effect as shown in Fig.~\ref{fig:dep}) and can evaporates.
   For model \texttt{h159}, this leads to the decrease of the averaged star formation 
   rate with respect to the static model (\emsta). This effect is however not seen in model \texttt{h070} for which
   the star formation rate of the \emrho\,model exceeds the one with model \emsta. While a deeper analysis would be needed here, we interpret this difference
   by the deeper potential well of model \texttt{h070} compared to model \texttt{h159} at $z=z_\textrm{ext}$. 
   In this case, the gravitational confinement of the gas dominates over the pressure one.
   Finally, when the hot gas temperature scales with respect to the gas density (\emrhoT, red curves), at the infall time, the thermal pressure of the hot 
   gas is no longer present to confine the dwarf gas, as shown by its very low $\beta_\textrm{RP}$ in Fig.~\ref{fig:beta}. Despite its increase at later
   time ($t > 6\,\rm{Gyr}$), the ram pressure stripping no longer impact the dwarf, as its cold gas already evaporated.
   In this model, both dwarf exhibit a star formation rate comparable to the isolated case sharing the same final stellar mass.
   As discussed in Sec.~\ref{sec:sfr_impact}, the small differences at time $t < 2$ Gyr is due to the method used to compute the stellar content of the dwarf.

    \subsubsection{Impact on the final dwarf properties}

\begin{figure*}
     \begin{center}
       \includegraphics[width=1.1\textwidth]{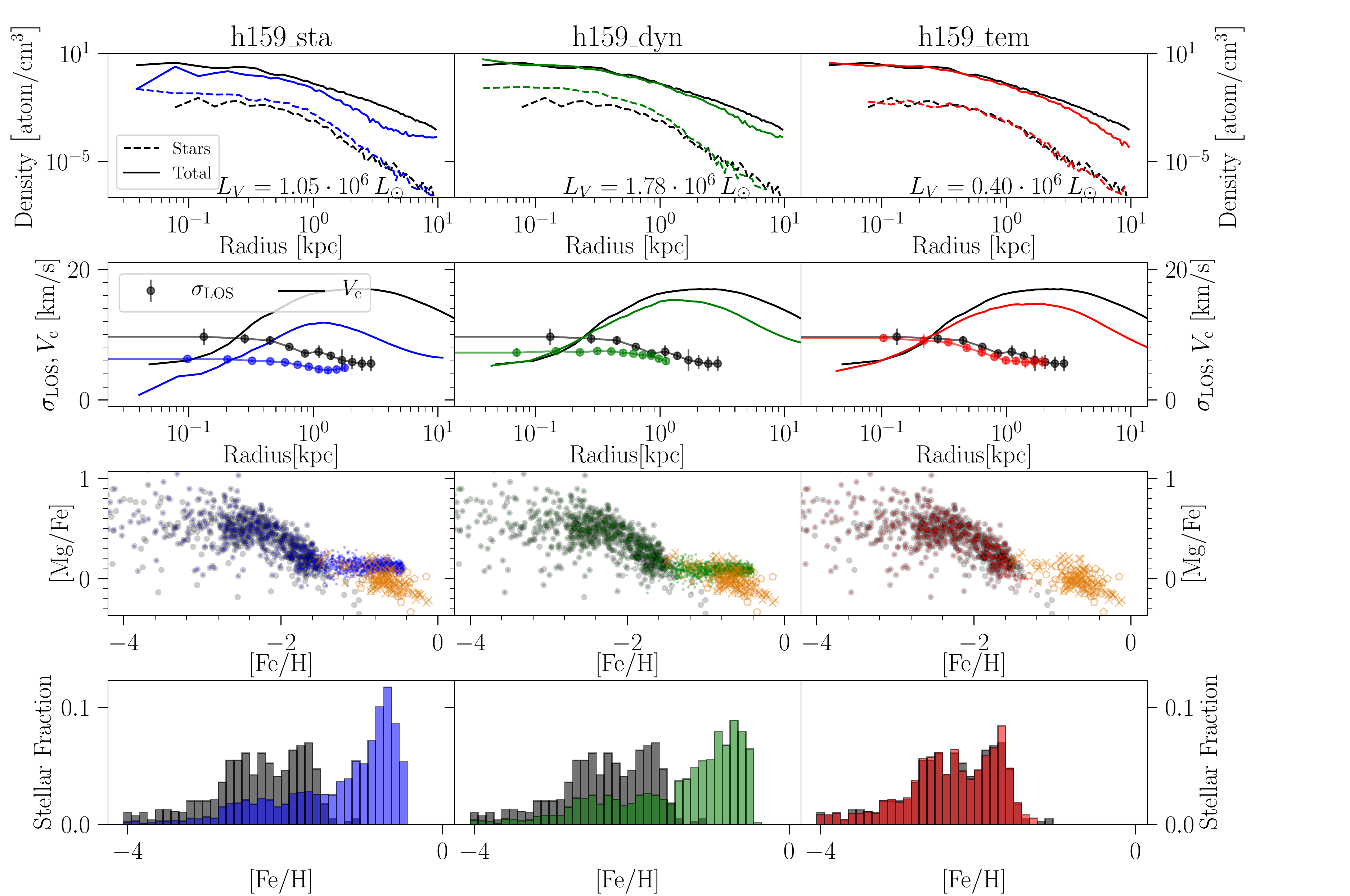}
       \caption{Properties of the different simulations in a moving box.
         From left to right, the simulations are in blue \texttt{h159\_sta}, \texttt{h159\_dyn} and \texttt{h159\_tem} and in red \texttt{h159\_iso}.
         In the first line, the density profile is shown for the total mass (straight lines) and the stellar mass (dashed lines).
         In the second line, the circular velocity and line of sight velocity dispersion are shown.
         In the third line, the stellar [Mg/Fe] vs [Fe/H] distribution is shown.
         The orange crosses (pentagons) are observations of the LMC bar (inner disc) \citep{van_der_swaelmen_chemical_2013}.
         In the last line, the metallicity distribution is shown.
         % /home/epfl/lhausamm/archive/paper2018/plot_article/real_cond/all_figures.py
         \label{fig:all3}
         }
     \end{center}
   \end{figure*}

\begin{figure*}
     \begin{center}
       \includegraphics[width=1.1\textwidth]{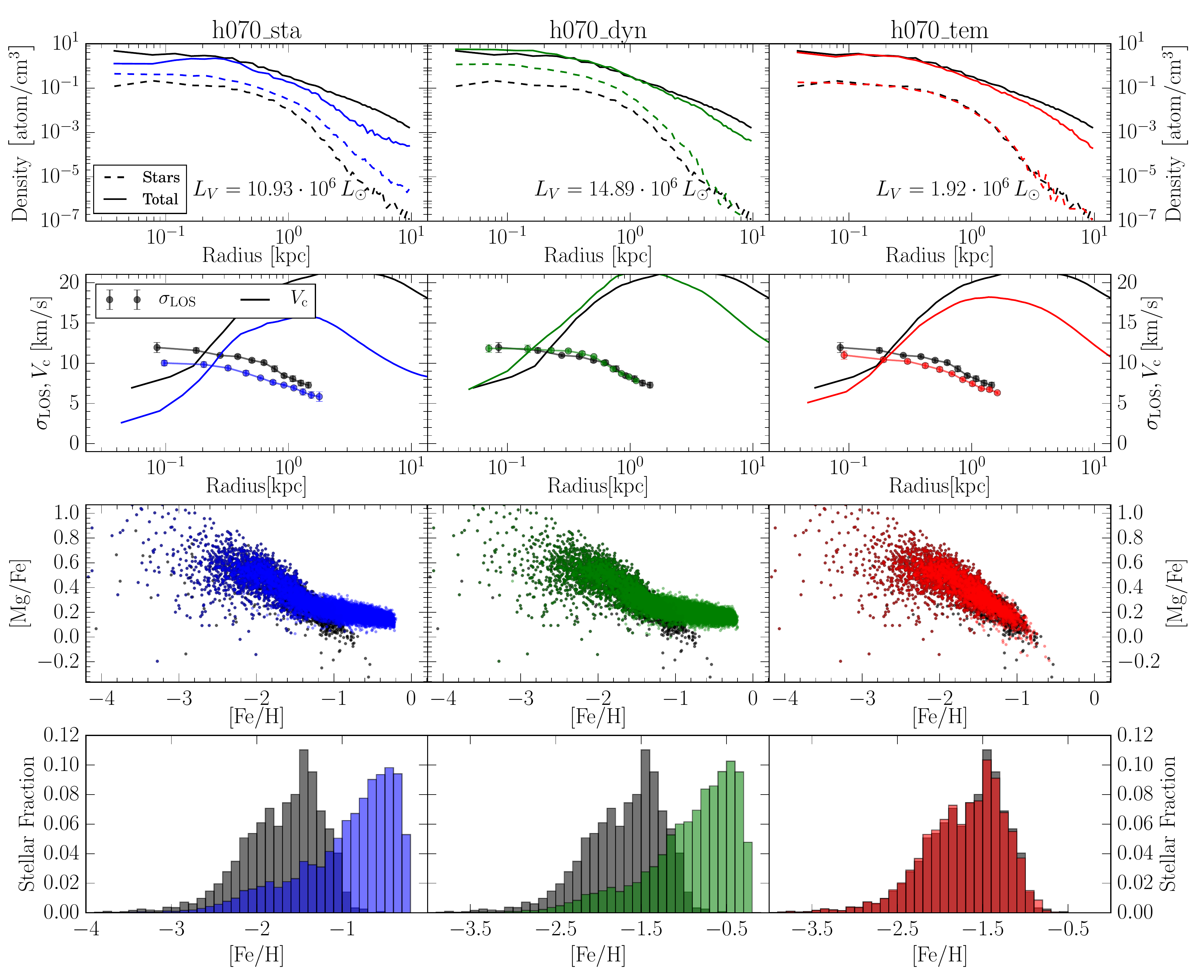}
       \caption{Properties of the different simulations in a moving box.
         From left to right, the simulations correspond to \texttt{h070\_sta}(blue), \texttt{h070\_dyn}(green) and \texttt{h070\_tem}(red)
         and are compared to the isolated model \texttt{h070\_iso} in grey.
         In the first line, the density profile is shown for the total mass (straight lines) and the stellar mass (dashed lines).
         In the second line, the circular velocity and line of sight velocity dispersion are shown.
         In the third line, the stellar [Mg/Fe] vs [Fe/H] distribution is shown.
         The orange crosses (pentagons) are observations of the LMC bar (inner disc) \citep{van_der_swaelmen_chemical_2013}.
         In the last line, the metallicity distribution is shown.
         % /home/epfl/lhausamm/archive/paper2018/plot_article/real_cond/all_figures.py
         \label{fig:all3_70}
         }
     \end{center}
   \end{figure*}

    Figure~\ref{fig:all3} and \ref{fig:all3_70} present the final properties of the three interacting models of the dwarf \texttt{h159} and \texttt{h070}  with the Milky Way, 
    compared to their reference model in isolation.
   
    In those two figures, the first row displays the stellar density profile (dashed line) along with the total density profile including the dark halo (continuous line).
    While none of the the interacting models are destroyed, they all show clear sign of stripping at radius larger than about $1~\rm{kpc}$, where the total density profiles
    drop compared to the isolated case. With four passages at the pericentre, \texttt{h159\_sta} and \texttt{h070\_sta} (\emsta) are the most affected ones.
    They also see their total density profiles reduced up to 30\% in the inner regions.
    However, due to their extended star formation rates, both models \texttt{\_sta} and \texttt{\_rho} exhibit a denser stellar density profile. 
    On the contrary, with its quenched star formation history, the stellar profile of \texttt{h159\_tem} is similar to the isolated case.
    
    The tidal stripping also impacts the stellar line of sight velocity dispersion profile showed in the second row. In five of the six interacting models, 
    the velocity dispersion is lower than in the isolated case, up to $5\,\rm{km/s}$ for the \texttt{h159\_sta} model.  
    This decrease reflects the adiabatic decompression after the removal of the outer dark halo, also responsible of the reduction of the circular velocity.
    It is worth noting that this stripping could help reproduce the low velocity dispersion (down to $5\,\rm{km/s}$) observed in six Andromeda 
    galaxies and difficult to reproduce in isolated models \citep{revaz_pushing_2018}.
    Only model \texttt{h070\_dyn} sees its velocity dispersion and circular velocity increase in the central regions. This reflects its larger
    stellar content owing to its higher star formation rate.
    
    The third and fourth rows of Fig.~\ref{fig:all3} and \ref{fig:all3_70} compare the final chemical properties of the simulated dwarfs.
    The third row displays the abundance ratio of $\alpha$-elements, traced here by the magnesium as a 
    function of [Fe/H]. 
    Because the dwarf galaxies enter their host halo at $t\cong 2\,\rm{Gyr}$, the old metal poor stellar population ($\rm{[Fe/H]}\lesssim-1.5$)
    is not affected by the interaction. The stellar [Mg/Fe] vs [Fe/H] distribution is characterized by a plateau at very low metallicity ($\rm{[Fe/H]}\lesssim-2.5$) followed by
    a decrease of [Mg/Fe], corresponding to the period where SNeIa yields dominates overs the SNeII, due to the drop of the star formation rate.    
    In both the \texttt{\_sta} and \texttt{\_rho} models, the interaction with the hot gas halo leads to the extension of the star formation period.
    Therefore a new set of SNeII explode at a continuous rate and produce a constant injection of $\alpha$-elements, quickly locked into new formed stars. 
    This results into the formation of a plateau in [Mg/Fe] extending from $\rm{[Fe/H]}\cong -1.5$ to $\rm{[Fe/H]}\cong -0.5$ for model \texttt{h159}
    and  $\rm{[Fe/H]}\cong -1.4$ to $\rm{[Fe/H]}\cong -0.2$ for model \texttt{h070}.
    The large amount of stars formed at those metallicities are responsible of a peak in the metallicity distribution function shown in the fourth row.
    This peak is strongly shifted towards higher metallicities compared to the isolated case.
    While [Mg/Fe] plateau have been observed for metal rich ([Fe/H] $\gtrapprox$ -0.6) stellar population in 
    Sagitarius \citep{hasselquist_apogee_2017,carlin_chemical_2018}, Fornax and LMC \citep{van_der_swaelmen_chemical_2013}, and at a lower lever for Sculptor \citep[see Fig.~11 of][]{tolstoy_star-formation_2009}, 
    they are found at solar or sub-solar [Mg/Fe], much lower than the one obtained here.
    A similar plateau may be obtained, to a somewhat shorter extension, for the brightest dwarf models of \citep{revaz_pushing_2018}.
    While a dedicated study will be necessary, we claim that such plateau could also be obtained if \texttt{h070} would have entered its host halo 
    at about $4-5\,\rm{Gyr}$, the time needed to decrease [Mg/Fe] down to solar values, as shown in Fig.~\ref{fig:all3_70}. 
    
    Finally, as their star formation history are similar to the isolated case, model \texttt{h159\_tem} and \texttt{h070\_tem} (\emrhoT) do not display any significant difference in 
    their final chemical properties.

   %------------------------------------------------------------%
   %------------------------- Section --------------------------%
   %------------------------------------------------------------%

   \section{Discussion}\label{sec:discussion}

   Contrary to the widespread idea that local group dwarf spheroidal galaxies are easily quenched and devoid of gas due to 
   the ram pressure stripping induced by its hot host halo, our simulations reveal a more complex picture.
   Both our wind tunnel and moving box simulations show that, while the hot gas of the dwarf is quickly ram pressure stripped,
   its cold and clumpy gas is not. On the contrary, due to the confinement of this gas by the thermal pressure of the hot halo gas
   which hamper the evaporation of the dwarf gas, the mass fraction of this cold phase can stay much above the one observed in 
   the isolated case. 
   Consequently depending on the orbital parameters of the dwarf, its infall time and the temperature of the host galaxy hot halo, 
   the star formation of the dwarf may be extended over several Gyr or even heavily sustained up to the present time.
   
   %In this study, our moving box simulations relayed only on one single orbit. While this choice was motivated 
   %by the difficulty of exploring a large parameter space as well as our constraint on the rather distant
   %perigalacticon imposed by the moving box technique, it is worth asking if our results are general enough to lead to 
   %strong conclusions. Despite not explicitly taking into account the gravitational effect of the host galaxy, our wind tunnel
   %simulations explore a much larger parameter space and are, in this sense complementary to the moving box models. We argue
   %that running additional models with varied orbits will then only confirm the trends obtained by the wind tunnel simulations.

   \subsection{Comparison with other simulations}
   
   Numerous publications have been dedicated to the study of ram pressure stripping of galaxies, including our own.
   
   Some of them concluded to the efficient stripping of gas implying the truncation or dampening of star formation
   \citep{mayer_simultaneous_2006,yozin_transformation_2015,fillingham_under_2016,emerick_gas_2016,steinhauser_simulations_2016}.
   On the contrary, others concluded to the enhancement or reignition of the star formation
   \citep{bekki_starbursts_2003,kronberger_influence_2008,kapferer_effect_2009,nichols_post-infall_2015,salem_ram_2015,henderson_significant_2016,wright_reignition_2019}.
   While a bunch of studies concluded that ram pressure may lead to both effects\citep{bahe_competition_2012,bekki_galactic_2014} or no major effect \citep{williamson_chemodynamics_2018}.
   Those differences suggest that conclusions reached could strongly depend on the numerical methods used as well as the way the baryonic physics is implemented.
   Indeed, in those studies, a variety of hydrodynamical methods have been used. Those simulations relie on
   lagrangian SPH methods, with or without modern pressure-entropy formulation, eulerian methods with or without adaptive mesh refinement, 
   or hybrid ones like moving-mesh methods.  
   They differ by specific implementations of the ISM treatment, like radiative gas cooling below $10^4\,\rm{k}$,
   external UV-background heating, hydrogen self-shielding against the UV-ionizing photons or magnetic field.   
   Finally, they covers a large resolution range.
   
   %Indeed, for the published works giving enough information necessary to compute $\beta_\textrm{RP}$, exploring comparable thermal to ram pressure regimes,
   %a variety of numerical methods (particle-based approach as SPH, with or without modern pressure-entropy formation, grid-based methods with or without adaptive mesh refinement, or hybrid methods like moving-mesh methods),  resolutions, stellar feedback, and specific ISM treatment (like radiative gas cooling below $10^4\,\rm{k}$,
   %external UV-background heating or hydrogen self-shielding against the UV-ionizing photons) are used.
   
   We discuss hereafter differences in our approach compared to other works that may lead to discrepancies, 
   but also review works in different contexts that support our conclusions. Finally, will discuss our 
   results in an observational context.
  
   \subsubsection{Stripping in dwarf galaxies}
   
   In a seminal paper, \citet{mayer_simultaneous_2006} showed that ram pressure stripping was efficient at completely removing the dwarf satellite ISM as long
   as they have a sufficiently low pericentre. 
   %This result is in strong disagreement with our conclusions, where none of our models are able to get entirely rid of
   %their gas, as cold gas remains. 
   %We argue that the origin of this disagreement lines in the difference of the ISM treatment.
   However in their approach, the gas is not allowed to radiatively cool below $10^4\,\rm{K}$. Under those conditions, the gas stays in a 
   warm-hot and diffuse phase which is indeed easy to strip, as we demonstrated in
   Section~\ref{sec:hot_gas_stripping}. When the gas is allowed to cool down to lower temperature, it becomes clumpy (see Fig.~4 of \citet{revaz_pushing_2018}) and 
   exposes a smaller surface to the wind hampering an efficient momentum transfer between the wind and the cold gas. 
   %Consequently, the latter is harder to be stripped.
   Efficient satellite ram pressure stripping have also been recently mentioned by \citet{simpson_quenching_2018}, where the quenching of satellite star formation in 30 
   cosmological zoom simulations of Milky Way-like galaxies have been studied. In these simulations, up to 90\% of satellites with stellar mass equal to about 
   $10^6\,\rm{M}_\odot$ are quenched, with ram pressure stripping being identified to be the dominant acting mechanism. This is nicely illustrated
   in their Fig.~8. However, those simulations also reveal a lack of any cold and clumpy phase which would be difficult to strip. 
   In addition to a slightly lower resolution compared to ours, the absence of cold phase is the result of the stiff equation of state used, that represents a two-phase 
   medium in pressure equilibrium \citep{springel_cosmological_2003}. However, it prevents the gas to cool down to low temperatures.

   Recently, in high resolution simulation, \citet{emerick_gas_2016} studied the ram pressure stripping of Leo T-like galaxies, including the effect of supernovae, 
   and the presence of cold gas gas. However, they do not include a fully self-consistent star formation method, supernova
   being exploded at a location determined by a randomly sampled exponentially decreasing probability distribution centred on the galaxy.
   While concluding that the RP is unable to completely quench these type of galaxies in less than $2\,\rm{Gyr}$, they show a clear decrease of the cold gas, 
   contradicting our results. While being cooler and denser that the gas considered in \citet{mayer_simultaneous_2006} due to a temperature floor of $6\cdot 10^3\,\rm{K}$,
   as illustrated by their Fig.~3, it is nevertheless not as clumpy as the one considered in our work.  

   \subsubsection{Thermal pressure confinement}

   One of the key effect that prevent the cold gas to evaporate and help sustain the star formation 
   in our simulations is the thermal pressure confinement of hot ambient gas. 
   We show hereafter that this effect is not only specific to our simulations but has been observed in other contexts. 

   Relying on SPH N-body simulations, \citet{bekki_starbursts_2003} studied the hydrodynamical effects of the hot ICM on 
   a self-gravitating molecular gas in a spiral galaxy. They concluded that the high pressure of the Intra Cluster Medium (ICM) can trigger the collapse of 
   molecular clouds leading to a burst of star formation. 
   Along the same line, \citet{kronberger_influence_2008} mentioned that in their models, the star formation rate is significantly enhanced by the ram-pressure 
   effect (up to a factor of 3) when a disk galaxy move through an idealized ICM.
   Similarly, ram pressure can favour H$_2$ formation \citep{henderson_significant_2016}, indirectly boosting the formation of stars.
   
   \citet{mulchaey_hot_2010} observed the hot halo surrounding galaxies and found a deficit of X-ray in comparison to the K-band luminosity for field galaxies,
   when compared to the galaxies in groups or cluster. They interpreted this results as the possibility that, contrary to field galaxies than can loose gas
   by supernova-driven winds, galaxies in groups or clusters see this outflowing material being pressure confined, preventing it to leave the galaxy halo.

   In a more quantitative way, using the GIMIC simulations, \citet{bahe_competition_2012} explored the pressure confinement by studying its effect on normal galaxies falling 
   in groups or cluster, directly computing the coefficient $\beta_\textrm{RP}$. In their simulations, they found 16\% of their galaxies to be dominated by thermal pressure. 
   %($\beta_\textrm{RP}$>1) with however a minor impact on the retention of hot gas.

   Sign of star formation increase due to confinement pressure have been also mentioned for simulation at a dwarf 
   scale. This effect has been described by \citet{nichols_post-infall_2015} in their dwarf spheroidal simulations with 
   however a star formation boost lower than the ones obtain in the present paper. While this study also relied on the 
   moving box technique, the physical prescriptions used where not comparable to the one used in the present study. The 
   simulations where run out of any cosmological context and neither UV-background nor hydrogen self-shielding where 
   considered.

   \citet{williamson_chemodynamics_2018} used a technique similar to our moving box and observed a thermal confinement.
   While the ram pressure has a negligible impact on the star formation, they showed that the outflows are confined and
   slightly increase the metallicity of the dwarf.

   \citet{wright_reignition_2019} observed that in their cosmological simulations, dwarf galaxies can re-ignite 
   star formation, following the complete quenching of the galaxy due to UV-background heating. This re-ignition 
   results from the compression of remaining hot gas in the dwarf halo, following an interaction with streams of gas in 
   the Inter Galactic Medium (IGM). Those gas streams being either due to cosmic filaments or resulting from nearby galaxy mergers.
   This mechanism is particularly efficient when the ram pressure is low compared to the thermal pressure (high $\beta_\textrm{RP}$), 
   which corroborates with our own results.

   \subsection{Comparison with observations}

   From the observational point of view, the idea that satellites galaxies have been ram pressured stripped is mainly 
   supported by the morphology-density relation observed in the Local Group 
   \citep{einasto_missing_1974,van_den_bergh_outer_1994,grcevich_hi_2010}. Quenched gas-poor spheroidals galaxies are 
   found in the vicinity of their host galaxy ($R \lesssim 300\,\rm{kpc}$) while star forming gas-rich dwarf irregulars 
   are found at larger distances. At the exception of Leo I, Fornax and Carina that show a very recent quenching time\citep[see 
   for example][]{skillman_islands_2017} the majority of dSphs have been quenched at least $5\,\rm{Gyr}$ ago.
   
   % Leo T, 
   %At first sight, our results could appear to be in strong disagreement with this relation, as the gas of nearby 
   %satellites is expected to be pressure confined and should, according to our models, present either gas or signs of 
   %ongoing or recent star formation. Despite Leo I, Fornax and Carina showing a very recent quenching time \citep[see 
   %for example][]{skillman_islands_2017} the majority of dwarf have been quenched at later time. However, the reality could well 
   %be more complex.
   
   We point out that recent observational facts suggest that the morphology-density relation may not be 
   universal. Indeed, spectroscopic observations of satellites galaxies around the NGC 4258 group showed that the 
   majority of the 16 detected probable and possible satellites, lying within a $250\,\rm{kpc}$, with a V-band 
   magnitude down to -12, appears to be blue star-forming irregular galaxies in the SDSS image \citep{spencer_survey_2014}. 
   This is in strong contrast with the observations of the Local Group.
   
   More recently, the SAGA survey \citep{geha_saga_2017} observed satellites companions around eight Milky Ways analogues, 
   with luminosities down to the one of Leo I ($M_\textrm{r} < -12.3$), equivalent to about $M_\star = 10^6$ M$_\odot$ 
   for star forming galaxies and $M_\star = 10^7$ M$_\odot$ for quenched galaxies. They found that among the 27 dwarf 
   detected, the majority, 26 galaxies are star forming. This results points towards a less efficient quenching in 
   those galaxies, compared the Milky Way.
    
   The star formation rate of our models is strongly dependent on the infall time of the satellite relatively to the 
   time when the galaxy halo is sufficiently hot and dense. As shown in Sect.~\ref{sec:effect_of_mw_model} when 
   the secular increase of the density and its temperature are taken into account, the evolution of the dwarfs entering 
   the halo before a redshift of 2.4 are hardly different from those of their isolated counterparts. 
   In that case the thermal pressure is unable to confine the gas of the satellites. This possibly 
   could reflect that the  different satellite population observed between the Milky Way and M31 and the ones of the 
   SAGA survey could simply reflect a difference in the assembly history of the host galaxies.

   \subsection{Additional potential heating/cooling sources}

   As mentioned in section \ref{sec:gear}, our current cooling implementation does not include $H_2$.
   Adding this efficient coolant will increase the fragmentation of the gas,
   making it even more clumpy, strengthening our results.

   It is worth mentioning that increasing the heating of the dwarf ISM could obviously help in quenching the star formation 
   by ejecting more gas.
   Boosting the stellar feedback is not a viable solution as it would fail to 
   reproduce the chemical observed properties of dwarf galaxies \citep{revaz_pushing_2018}.

   Another possible heating source is the thermal conduction between the MW's hot halo and the dwarf's cold gas.
   \citet{cowie_evaporation_1977} and \citet{mckee_evaporation_1977} developed an analytical model for the evaporation of an
   isolated spherical cloud in a hot gas.
   They considered both classical (electrons' mean free path smaller than the cloud size) and saturated thermal conduction (electrons' mean free path comparable to the cloud).
   Their analytical model shows that our dwarfs do not enter any saturated regime and are only marginally dominated by radiation loss.
   While detailed numerical simulations would be necessary to provide a conclusive answer, this first approximation predicts an evaporation over several Gyr.

   Finally, considering the high UV-flux emitted by the proto-host Galaxy \citep{van_den_bergh_outer_1994} or 
   the potential strong impact of an AGN could help in removing the remaining confined gas.

   %% \subsection{Additional tests}
   
   %% As our models involve a rather complex treatment of the baryonic physics, it is worth asking if our 
   %% results could suffer from numerical issues. We recall here that the SPH hydro-solver implemented in \texttt{GEAR} 
   %% follows the pressure-entropy SPH formulation \citep{hopkins_general_2013} which ensures the correct treatment of 
   %% fluid mixing instabilities. Without this essential feature, even the hot halo gas would have been difficult to 
   %% strip, as demonstrated by former tests.
   
   %% However, in order to double check the validity of our implementation, we performed an additional test which consist 
   %% in simulating a Milky Way galaxy and its interaction with an ICM. In those conditions, observations 
   %% clearly shows signs of cold gas stripping that should be capture in our simulations. The set-up and results of those 
   %% tests are described in Appendix~\ref{an:agora}. They demonstrate that when the confinement pressure is very low with 
   %% respect with the ram pressure ($\beta_{\rm{RP}}\ll 1$), the cold gas phase is ram pressure striped, as expected from 
   %% a theoretical point of view.

   %------------------------------------------------------------%
   %------------------------- Section --------------------------%
   %------------------------------------------------------------%
   \section{Conclusions}\label{sec:conclusion}

   We have presented high resolution \texttt{GEAR}-simulations of the
   interaction of dwarf spheroidal galaxies formed in a cosmological
   $\Lambda$CDM context with a Milky Way-like galaxy.  We first ran a large set
   of wind tunnel simulations focusing on the hydrodynamical
   interaction between the dwarf system and the MW hot halo gas. We varied the
   wind parameters, which describe the velocity at which the dwarf enters the
   hot halo and orbits around the central galaxy, as well as the density and the
   temperature of the host halo gas. This allowed us to investigate how the ISM
   of the dwarf satellite was modified and to infer how its cold and hot gas
   phases could be ram pressure stripped. In a second step, we performed a set
   of moving box simulations that added the gravitational tidal
   interactions to the hydrodynamical ones. We also included the 
   variation of the density and temperature of the hot halo all along 
   the dwarf orbit as well as their increase due to the secular growth of the Milky Way.

   %While our main goal was to demonstrate that isolated dwarf galaxies characterized by an extended star formation 
   %history are quickly quenched by ram pressure stripping when entering the hot gas halo of their host galaxy, we 
   %discovered a much more complex picture questioning well established idea.

   The conclusions we reach are significantly different from those of previous works. 
   Indeed, it turns out that including the hydrogen-self shielding
   that allows the gas to cool much below $10^4\,\rm{K}$, leading to a multiphase ISM, absent in most of the previous 
   studies, is essential to capture the effect of the ram pressure stripping and its impact on the dwarf star formation 
   history.

   Our results can be summarized as follows:
   
   \begin{itemize}

   \item While the hot and diffuse gas phase of the dwarf ($T>1000\,\rm{K}$) is efficiently 
     and quickly stripped by the ram pressure induced by the gas of its host halo,
     the cold, star forming and clumpy gas phase ($T<1000\,\rm{K}$) is not necessarily.     
     The efficiency of the stripping of this cold gas 
     depends on the ratio between the thermal pressure and the ram pressure both exerted on 
     the dwarf by the hot halo gas. When the thermal pressure is high, the cold gas is confined
     and its stripping is slowed down.

    \item As a consequence of the above, the infall time of a dwarf galaxy plays
      a decisive role in the evolution of the dwarf satellites.  If the
      interaction between the host galaxy and its satellite begins when the
      thermal pressure is low, that is the host halo is not
      sufficiently dense or hot, then, the evolution of the dwarf will be
      essentially the same as in isolation. The cold ISM will evaporate due to the UV-background 
      heating and star formation will be quenched.
      On the contrary, the cold ISM is confined and remains attached to the dwarf.

   \item The confinement of the cold gas in the dwarf satellite leads to an
     extension of its star formation history. While the same dwarf galaxy
     would see its star formation quenched due to the evaporation of the residual
     gas, its interaction with the Milky Way keeps the star
     formation rate roughly at the level it had when the dwarf entered the host
     halo. This translates into a higher final mean metallicity, by up to 1 dex
     in the examples presented in this study.  Because our model dwarf spheroidals
     enter the Milky-Way like galaxy at $\sim 2\,\rm{Gyr}$, their star formation rates
     have already significantly decreased, therefore
     the ejecta of the SNeIa explosion contribute significantly to the dwarf's ISM
     enrichment.
     Hence, both our details models display an extended low, although still super-solar,
     [$\alpha$/Fe] tail. 
     A solar or sub-solar plateau similar to the Fornax or Sagittarius dwarf galaxy could 
     be obtained if the dwarf enters the hot halo of its host galaxy at later time, where 
     the [$\alpha$/Fe] decreased to lower values.
     Firm conclusion on this point would require a dedicated and thorough
     investigation.  

   \end{itemize}

   Ram-pressure and tidal interactions do not seem sufficient to explain by
   themselves the morphology-density relation observed in the Local Group. It
   would require very specific conditions, either a very late entry of the
   closest dSphs in the halo of the Milky Way, or a very early accretion before
   the end of the Galaxy mass assembly. 
   %{\bf mais est-ce compatible avec les
   %  proper motions, positions, orbites sans fatal infall comme Sag ?? etc
   %  ???}. 
   Other processes might play a role, such as the heating by the UV-flux of
   the Milky Way itself.
   
   Star forming satellites have been found around other Milky Way
   analogues or in groups \citep{spencer_survey_2014,geha_saga_2017}. As the
   effect on the hot host halo strongly depends on the infall time of the
   satellite galaxy, the different satellite populations observed between the
   Milky Way and M31 (dominance of quenched gas-poor galaxies) and the ones of
   the SAGA survey (star forming galaxies) could potentially reflect a
   difference in the assembly history of the host galaxies.

%------------------------------------------------------------%
%------------------------- Acknowledgements -----------------%
%------------------------------------------------------------%  

\begin{acknowledgements}

We are indebted to the International Space Science Institute (ISSI),
Bern, Switzerland, for supporting and funding the international team
'First stars in dwarf galaxies'.  We are grateful to Matthew Nichols
to help us with the moving box technique.  We enjoyed discussions with
Matthieu Schaller, Romain Teyssier, Françoise Combes, Allessandro
Lupi, Andrew Emerick.  This work was supported by the Swiss Federal
Institute of Technology in Lausanne (EPFL) through the use of the
facilities of its Scientific IT and Application Support Center
(SCITAS). The simulations presented here were run on the Deneb
clusters.
The data reduction and galaxy maps have been performed
using the parallelized Python \texttt{pNbody} package
(\texttt{http://lastro.epfl.ch/projects/pNbody/}).
We are grateful to the \texttt{Numpy} \citep{oliphant_guide_2015},  \texttt{Matplotlib} \citep{thomas_a_caswell_matplotlib/matplotlib_2018}
\texttt{SciPy} \citep{jones_scipy:_2001} and IPython \citep{perez_ipython:_2007}
teams for providing the scientific community with essential python tools.
\end{acknowledgements}

\bibliographystyle{aa}
\bibliography{dSph}

\appendix

\section{Additional wind tunnel and moving box simulations}

\begin{table*}
  \begin{center}
    \caption{List of additional wind tunnel simulations performed.
      The first three parameters are for the wind.
      The four last columns are the properties (total mass, luminosity, cold gas mass and hot gas mass) of the galaxies at the end of the simulation.
    }
    \begin{tabular}{l|ccccccc}
      \hline\hline
      Model                & $\rho_w$ [$10^{-5}$ atom/cm$^3$] & $u_w$ [km/s] & $T_w$ [$10^6$K]& $M_{200}$ [$10^8$ M$_\odot$] & $L_\textrm{V}$ [$10^6$ L$_\odot$] & Cold Gas [$10^6$ M$_\odot$] & Hot Gas [$10^6$ M$_\odot$] \\ \hline
	 \texttt{h050}  & 1.277 & 77 & 2.00 & 21.8 & 59.10 & 42.42 & 74.3 \\
	 \texttt{h050}  & 1.277 & 100 & 2.00 & 21.5 & 52.85 & 34.14 & 67.9 \\
	 \texttt{h050}  & 1.277 & 130 & 2.00 & 21.0 & 44.49 & 27.14 & 64.5 \\
	 \texttt{h050}  & 1.277 & 169 & 2.00 & 20.3 & 34.40 & 22.96 & 56.6 \\
	 \texttt{h050}  & 2.160 & 77 & 2.00 & 22.0 & 57.06 & 38.58 & 93.1 \\
	 \texttt{h050}  & 2.160 & 100 & 2.00 & 21.6 & 49.54 & 39.39 & 85.9 \\
	 \texttt{h050}  & 2.160 & 130 & 2.00 & 20.9 & 39.83 & 26.27 & 83.9 \\
	 \texttt{h050}  & 2.160 & 169 & 2.00 & 20.3 & 29.24 & 13.27 & 85.5 \\
	 \texttt{h050}  & 2.810 & 77 & 2.00 & 22.2 & 55.03 & 42.15 & 116.0 \\
	 \texttt{h050}  & 2.810 & 100 & 2.00 & 21.7 & 47.64 & 37.86 & 106.1 \\
	 \texttt{h050}  & 2.810 & 130 & 2.00 & 20.8 & 35.38 & 24.37 & 99.2 \\
	 \texttt{h050}  & 2.810 & 169 & 2.00 & 20.3 & 25.98 & 15.47 & 99.3 \\
	 \texttt{h050}  & 1.660 & 77 & 2.00 & 21.8 & 56.74 & 35.09 & 88.2 \\
	 \texttt{h050}  & 1.660 & 100 & 2.00 & 21.6 & 51.61 & 31.76 & 81.9 \\
	 \texttt{h050}  & 1.660 & 130 & 2.00 & 21.1 & 43.20 & 26.47 & 73.8 \\
	 \texttt{h050}  & 1.660 & 169 & 2.00 & 20.3 & 31.82 & 19.97 & 66.3 \\
	 \texttt{h050}  & 1.277 & 169 & 1.30 & 20.0 & 27.80 & 22.31 & 54.1 \\
	 \texttt{h070}  & 1.277 & 77 & 2.00 & 13.0 & 19.53 & 16.70 & 34.7 \\
	 \texttt{h070}  & 1.277 & 130 & 2.00 & 12.6 & 13.30 & 11.35 & 28.4 \\
	 \texttt{h070}  & 1.277 & 169 & 2.00 & 12.5 & 11.69 & 9.17 & 28.7 \\
	 \texttt{h070}  & 1.277 & 100 & 2.00 & 12.8 & 15.78 & 13.31 & 30.4 \\
	 \texttt{h070}  & 2.160 & 77 & 2.00 & 13.1 & 17.80 & 15.20 & 46.9 \\
	 \texttt{h070}  & 2.160 & 130 & 2.00 & 12.8 & 12.95 & 11.32 & 43.1 \\
	 \texttt{h070}  & 2.160 & 169 & 2.00 & 12.7 & 11.32 & 8.15 & 45.2 \\
	 \texttt{h070}  & 2.160 & 100 & 2.00 & 12.9 & 14.78 & 12.03 & 45.4 \\
	 \texttt{h070}  & 2.810 & 77 & 2.00 & 13.2 & 17.55 & 15.57 & 60.1 \\
	 \texttt{h070}  & 2.810 & 130 & 2.00 & 12.8 & 12.52 & 10.09 & 55.2 \\
	 \texttt{h070}  & 2.810 & 169 & 2.00 & 12.7 & 10.82 & 9.11 & 55.9 \\
	 \texttt{h070}  & 2.810 & 100 & 2.00 & 13.0 & 14.79 & 8.43 & 60.5 \\
	 \texttt{h070}  & 1.660 & 77 & 2.00 & 13.0 & 18.87 & 8.43 & 48.9 \\
	 \texttt{h070}  & 1.660 & 130 & 2.00 & 12.7 & 12.86 & 10.86 & 35.0 \\
	 \texttt{h070}  & 1.660 & 169 & 2.00 & 12.6 & 11.72 & 8.64 & 35.5 \\
	 \texttt{h070}  & 1.660 & 100 & 2.00 & 12.8 & 14.90 & 14.56 & 34.4 \\
	 \texttt{h070}  & 1.277 & 169 & 0.40 & 12.1 & 5.21 & 1.93 & 23.6 \\
	 \texttt{h070}  & 1.277 & 169 & 0.60 & 12.2 & 6.57 & 5.18 & 23.9 \\
	 \texttt{h070}  & 1.277 & 169 & 1.30 & 12.5 & 10.42 & 8.98 & 27.5 \\
	 \texttt{h070}  & 1.277 & 200 & 0.40 & 12.1 & 4.93 & 1.07 & 22.9 \\
	 \texttt{h070}  & 1.277 & 200 & 0.60 & 12.2 & 6.00 & 3.05 & 23.8 \\
	 \texttt{h070}  & 1.660 & 220 & 0.20 & 12.1 & 4.25 & 0.00 & 28.5 \\
	 \texttt{h070}  & 1.660 & 220 & 0.30 & 12.1 & 4.66 & 0.00 & 28.7 \\
	 \texttt{h070}  & 1.660 & 250 & 0.20 & 12.1 & 4.36 & 0.00 & 28.6 \\
	 \texttt{h070}  & 1.660 & 250 & 0.30 & 12.1 & 4.51 & 0.00 & 28.4 \\
	 \texttt{h070}  & 1.660 & 300 & 0.30 & 12.1 & 4.72 & 0.00 & 28.6 \\
	 \texttt{h070}  & 1.660 & 400 & 0.30 & 12.1 & 4.89 & 0.00 & 28.6 \\
	 \texttt{h123}  & 2.160 & 30 & 0.20 & 6.7 & 0.14 & 0.00 & 19.7 \\
	 \texttt{h123}  & 2.160 & 30 & 0.40 & 6.7 & 0.14 & 0.00 & 22.2 \\
	 \texttt{h123}  & 2.160 & 30 & 0.60 & 6.9 & 0.33 & 2.81 & 37.9 \\
	 \texttt{h123}  & 2.160 & 30 & 1.30 & 7.0 & 1.44 & 4.74 & 43.4 \\
	 \texttt{h123}  & 2.160 & 30 & 1.54 & 7.1 & 2.33 & 6.10 & 47.1 \\
	 \texttt{h123}  & 2.160 & 30 & 2.00 & 7.1 & 2.35 & 5.26 & 44.6 \\
	 \texttt{h123}  & 2.160 & 30 & 3.39 & 7.2 & 4.34 & 7.34 & 50.0 \\
	 \texttt{h132}  & 1.277 & 169 & 1.30 & 14.1 & 10.64 & 10.58 & 29.7 \\
	 \texttt{h168}  & 2.810 & 30 & 3.39 & 10.1 & 7.70 & 8.28 & 85.8 \\
	 \texttt{h074}  & 1.277 & 169 & 1.30 & 5.8 & 0.73 & 0.00 & 10.6 \\
	 \hline\hline
    \end{tabular}
    \label{tab:addwt}
  \end{center}
\end{table*}

\begin{table*}
  \begin{center}
    \caption{List of additional moving box simulations performed.
      The Milky Way model is given with the model name following the same convention than in table \ref{table:real_sim}.
      The four last columns are the properties (total mass, luminosity, cold gas mass and hot gas mass) of the galaxies at the end of the simulation.
}
    \begin{tabular}{l|ccccc}
      \hline\hline
      Model                & $M_{200}$ [$10^8$ M$_\odot$] & $M_\star$ [$10^6$ M$_\odot$] & $L_\textrm{V}$ [$10^6$ L$_\odot$] & Cold Gas [$10^6$ M$_\odot$] & Hot Gas [$10^6$ M$_\odot$] \\ \hline
      \texttt{h050\_sta} & 8.8 & 117.34 & 17.63 & 14.10 & 103.2\\
      \texttt{h050\_iso} & 11.4 & 36.09 & 7.90 & 14.67 & 21.4\\
      \texttt{h123\_tem} & 3.4 & 2.78 & 0.14 & 0.00 & 2.8\\
      \texttt{h123\_iso} & 3.7 & 1.02 & 0.13 & 0.00 & 1.0\\
      \texttt{h132\_rho} & 6.1 & 77.33 & 17.85 & 12.44 & 64.9\\
      \texttt{h132\_sta} & 2.8 & 60.77 & 13.40 & 9.25 & 51.5\\
      \texttt{h132\_tem} & 5.0 & 24.52 & 9.08 & 11.24 & 13.3\\
      \texttt{h132\_iso} & 16.0 & 0.01 & 2.88 & 0.00 & 0.0\\
      \texttt{h168\_tem} & 8.5 & 4.54 & 1.09 & 0.00 & 4.5\\
      \texttt{h168\_iso} & 6.1 & 2.98 & 1.02 & 0.00 & 3.0\\
      \texttt{h074\_rho} & 2.6 & 34.01 & 2.57 & 3.84 & 30.2\\
      \texttt{h074\_sta} & 1.2 & 26.15 & 2.62 & 3.81 & 22.3\\
      \texttt{h074\_tem} & 2.4 & 9.41 & 2.34 & 3.64 & 5.8\\
      \texttt{h074\_iso} & 6.2 & 0.00 & 0.65 & 0.00 & 0.0\\
    \end{tabular}
    \label{tab:addmb}
  \end{center}
\end{table*}

%#################################################################
\end{document}